\documentclass[11pt,a4paper]{article}
\pdfoutput=1

\usepackage{epsfig}
\usepackage{amsmath}
\usepackage{amssymb}
\usepackage{bbm}
\usepackage{verbatim}
\usepackage{bm}
\usepackage{color}
\usepackage{dsfont}
\usepackage{caption}
\usepackage[subrefformat=parens,labelformat=parens]{subfig}
\usepackage{subfloat}
\usepackage{cancel}
\usepackage{cite}
\numberwithin{equation}{section}

\usepackage{booktabs}

\def\d2{W''}

\definecolor{lightcoral}{rgb}{0.94, 0.5, 0.5}
\definecolor{mahogany}{rgb}{0.75, 0.25, 0.0}
\definecolor{wildstrawberry}{rgb}{1.0, 0.26, 0.64}

\begin{document}

\begin{flushleft}
DESY 15-080\\
CPHT-RR019.0515\\
June 2015
\end{flushleft}

\vskip 1cm

\begin{center}
{\Large\bf Moduli backreaction and supersymmetry breaking \\ in string-inspired inflation models} 

\vskip 2cm

{Emilian Dudas$^{a,b}$, Clemens Wieck$^b$}\\[3mm]
{\it{
$^a$ CPhT, Ecole Polytechnique, 91128 Palaiseau Cedex, France \\
$^b$ Deutsches Elektronen-Synchrotron DESY, 22607 Hamburg, Germany\\
}}
\end{center}

\vskip 1cm

\begin{abstract}
\noindent 
We emphasize the importance of effects from heavy fields on supergravity models of inflation. We study, in particular, the backreaction of stabilizer fields and geometric moduli in the presence of supersymmetry breaking. Many effects do not decouple even if those fields are much heavier than the inflaton field. We apply our results to successful models of Starobinsky-like inflation and natural inflation. In most scenarios producing a plateau potential it proves difficult to retain the flatness of the potential after backreactions are taken into account. Some of them are incompatible with non-perturbative moduli stabilization. In natural inflation there exist a number of models which are not constrained by backreactions at all. In those cases the correction terms from heavy fields have the same inflaton-dependence as the uncorrected potential, so that inflation may be possible even for very large gravitino masses.
\end{abstract}

\thispagestyle{empty}

\newpage

\tableofcontents

%
%%
%%%
%%%%
%%%%%%%%%%%%%%%%%%%%%%%%%%%%%%%%%%%%%%%%%%%%%%%
%%%%%%%%%%%%%%%%%%%%%%%%%%%%%%%%%%%%%%%%%%%%%%%
%%%%%%%%%%%%%%				      %%%%%%%%%%%%%%%%%%%%%%%
%%%%%%%%%%%%%%         Introduction      %%%%%%%%%%%%%%%%%%%%%%%
%%%%%%%%%%%%%%				      %%%%%%%%%%%%%%%%%%%%%%%
%%%%%%%%%%%%%%%%%%%%%%%%%%%%%%%%%%%%%%%%%%%%%%%
%%%%%%%%%%%%%%%%%%%%%%%%%%%%%%%%%%%%%%%%%%%%%%%

\section{Introduction}

Observations of the cosmic microwave background (CMB) radiation favor the paradigm that the very early universe can be described by a phase of single-field slow-roll inflation \cite{Ade:2015tva,Ade:2015lrj}. This can be realized in theories where a single scalar field, the inflaton field, is lighter than all others so that its potential satisfies the inflationary slow-roll conditions.\footnote{We recommend \cite{Baumann:2014nda} as a review.} However, additional fields can influence the predictions of many inflation models even when they are heavier than the Hubble scale. We analyze such effects in some of the earliest and most successful inflation models and their supergravity embeddings. 

One of the first models of inflation was developed by A.~Starobinsky \cite{Starobinsky:1980te}. It has an exponentially flat scalar potential which is generated by a non-minimal coupling to gravity. In recent years many successful attempts have been made to embed similar scenarios in four-dimensional $\mathcal N = 1$ supergravity. The ones we wish to study, as representative examples, are the Cecotti model and its generalization \cite{Cecotti:1987sa,Kallosh:2013lkr}, the class of no-scale supergravity models proposed in \cite{Ellis:2013xoa}, and the Goncharov-Linde model \cite{Goncharov:1983mw,Goncharov:1985yu}.

Another successful scenario to produce $50 - 60$ $e$-folds of inflation, called natural inflation, was developed in \cite{Freese:1990rb}. It contains a scalar field with a discrete axionic shift symmetry, resulting in a periodic potential usually described by a cosine function. For the predictions of this setup to agree with observations the axion decay constant of the inflaton field must be super-Planckian. Axions are abundant in string theory, but super-Planckian decay constants are forbidden \cite{Banks:2003sx,Svrcek:2006yi}. Hence, a number of ways have been proposed to generate effectively large axion decay constants in string-derived and string-inspired setups. Among those we wish to study are aligned natural inflation \cite{Kim:2004rp} and axion monodromy inflation \cite{Silverstein:2008sg,McAllister:2008hb}.

Many of the references given above analyze possible embeddings of the respective inflation models in string theory. In that case consistent stabilization of all moduli is generically a concern. In most cases, to realize single-field inflation without large amounts of isocurvature fluctuations or non-Gaussianities all moduli and extra fields must be heavier than the Hubble scale.\footnote{Again, we recommend \cite{Baumann:2014nda} as a review and a comprehensive list of references on moduli stabilization.} Even if all moduli are stabilized supersymmetrically, meaning in our terminology that the masses of the moduli are independent of the gravitino mass in the vacuum,\footnote{Note that we only consider Minkowski vacua. Here the gravitino mass always parameterizes supersymmetry breaking.} the inflationary vacuum energy induces a backreaction of the moduli on the inflaton potential. This kind of moduli stabilization, called strong or supersymmetric moduli stabilization, can be achieved using a racetrack setup \cite{Kallosh:2004yh,Dudas:2012wi}, or via interactions with additional fields in the presence of anomalous U(1) symmetries \cite{Wieck:2014xxa}. The backreaction of moduli stabilized in this way has been studied in \cite{Buchmuller:2014vda} for generic F-term inflation models. It is suppressed by powers of $H/m_{T_i}$, the Hubble scale divided by the mass of the moduli. 
If some of the moduli break supersymmetry in the vacuum, the situation is more complicated. Examples of this kind of moduli stabilization include the setup of KKLT \cite{Kachru:2003aw}, the Large Volume Scenario \cite{Balasubramanian:2005zx,Conlon:2005ki}, and K\"ahler Uplifting \cite{Balasubramanian:2004uy,vonGersdorff:2005bf,Berg:2005yu,Westphal:2006tn}. The interplay of such moduli with inflation has been studied in many instances. Here, we follow and extend the work of \cite{Buchmuller:2015oma} which is concerned with the backreaction in chaotic inflation with a quadratic potential. We demonstrate that the non-decoupling effects found in \cite{Buchmuller:2015oma} arise more generally and may have severe implications for both plateau-like inflation and natural inflation.\footnote{For a different approach involving the couplings between heavy and light superfields, cf.~\cite{Choudhury:2014sxa}.}

In addition to moduli fields from string compactifications there may be other heavy fields which cause relevant backreactions on the inflaton potential. In particular, some of the references given above make use of so-called stabilizer fields. It was shown in \cite{Buchmuller:2014pla} that the interplay between such stabilizer fields and supersymmetry breaking is non-trivial. Hence, in models with a stabilizer field we are concerned with the backreaction of the latter, which generically becomes important when the scale of supersymmetry breaking is comparable to the Hubble scale. Similar as for the backreaction of moduli fields, we find that plateau models are more susceptible to the effects of integrating out the stabilizer field. This is because the exponential flatness of the inflaton potential is easily spoiled by additional sectors in the theory.

This paper is organized as follows. In Sec.~2 we present a general discussion of the backreactions induced by stabilizer fields and supersymmetry-breaking moduli. We illustrate the results of this discussion in the following sections. In Sec.~3 we focus on supergravity models with stabilizer fields, treating plateau models and natural inflation models separately. The comparison between these two classes of models is particularly instructive. Sec.~4 is devoted to models without stabilizer fields, and the backreaction of heavy moduli, in particular K\"ahler moduli. Again we investigate plateau models first, and then natural inflation models. Finally, we give a short conclusion in Sec.~5.

%
%%
%%%
%%%%
%%%%%%%%%%%%%%%%%%%%%%%%%%%%%%%%%%%%%%%%%%%%%%%
%%%%%%%%%%%%%%%%%%%%%%%%%%%%%%%%%%%%%%%%%%%%%%%
%%%%%%%%%%%%%%				    %%%%%%%%%%%%%%%%%%%%%%%
%%%%%%%%%%%%%%         Section 2        %%%%%%%%%%%%%%%%%%%%%%%
%%%%%%%%%%%%%%				    %%%%%%%%%%%%%%%%%%%%%%%
%%%%%%%%%%%%%%%%%%%%%%%%%%%%%%%%%%%%%%%%%%%%%%%
%%%%%%%%%%%%%%%%%%%%%%%%%%%%%%%%%%%%%%%%%%%%%%%

\section{General remarks}
\label{sec:General}

Extra fields in theories describing inflation can have effects on the inflationary dynamics even if they are much heavier than the Hubble scale. In string-derived or string-inspired models these fields can be, for example, geometric moduli fields or so-called stabilizer fields. In the following we treat these two cases separately. We integrate out the heavy fields and compute their backreaction in the form of inflaton-dependent corrections to the scalar potential.\footnote{As in \cite{Buchmuller:2015oma,Buchmuller:2014pla} we consider setups in which the heavy fields trace their inflaton-dependent minima instantaneously and adiabatically. In this sense our approach is different from the one taken in \cite{Achucarro:2010jv,Achucarro:2010da,Achucarro:2012sm,Achucarro:2012yr,Achucarro:2015bra,Achucarro:2015rfa}, where the authors consider sharp turns of inflationary trajectories in the valleys of heavy fields. Although our models generically do not exhibit such features, it may be interesting to study a combination with the effects that we find.}

%
%%
%%%
%%%%
%%%%%%%%%%%%%%%%%%%%%%%%%%%%%%%%%%%%%%%%%%%%%%%
%%%%%%%%%%%%%%%%%%%%%%%%%%%%%%%%%%%%%%%%%%%%%%%
\subsection{Backreaction of heavy stabilizer fields}
\label{sec:GenStab}

A large class of supergravity models with a stabilizer field $S$ can be defined by the following effective Lagrangian,
\begin{align}\label{eq:Kahleransatz1}
K &= K (\Phi + {\overline \Phi}, S, \bar S, X, \bar X, T_\alpha, \overline T_\alpha) \, , \\
W &= M S f (\Phi) + W_1 (X, T_\alpha) \,, \label{g1}
\end{align}
where $f$ is a holomorphic function. Usually the inflaton field is proportional to the imaginary part of $\Phi$, which is protected by an axionic shift symmetry. The field $X$ is responsible for supersymmetry breaking, i.e., for an uplift of the post-inflationary vacuum to Minkowski or de Sitter spacetime. The fields $T_\alpha$ are additional degrees of freedom like moduli fields or the axio-dilaton whose effects we neglect for the moment. $M$ is a mass scale which sets the energy scale of inflation. The K\"ahler potential is usually chosen such that, in the absence of the supersymmetry-breaking piece $W_1$, $S$ is stabilized at the origin of field space. In this case the scalar potential on the inflationary trajectory becomes 
\begin{align}
V_\text{inf} \sim M^2 |f(\Phi)|^2\,.
\end{align}
In most cases $f(\Phi)$ vanishes after inflation so that the true vacuum is supersymmetric and Minkowski. Once $W_1$ breaks supersymmetry, however, there is a mixing between the stabilizer field and the inflaton field. Qualitatively it is of the form
\begin{align}
V_\text{soft} \sim  m_{3/2} \left[\text{Re}\,S f_1(\varphi) + \text{Im}\, S f_2(\varphi) \right]\,,
\end{align}
with $m_{3/2} = e^{K/2}W$ evaluated in the true vacuum.\footnote{Since $\langle S \rangle = 0$ during and after inflation, usually $m_{3/2} \propto \langle W_1 \rangle$.} $f_1$ and $f_2$ are model-dependent functions of the canonically normalized inflaton field $\varphi$. This mixing term forces the stabilizer field to track the evolution of the inflaton during inflation even if $S$ is much heavier than the inflaton, which is the case we consider. Through a second-order expansion in $\text{Re}\, S$ and $\text{Im}\, S$ one can find the inflaton-dependent minimum of the stabilizer field. It reads 
\begin{align}
\text{Re}\,S = -  m_{3/2} \frac{f_1 (\varphi)}{M_S^2 (\varphi)} \, , \qquad
\text{Im}\,S  = -  m_{3/2} \frac{f_2 (\varphi)}{M_S^2 (\varphi)} \, , \label{g2}
\end{align}
where $M_S^2 (\varphi)$ denotes the squared mass of the stabilizer field during inflation. Inserting this result in the scalar potential generates a backreaction term, i.e., the inflaton potential becomes
\begin{align}\label{g3}
V(\varphi) = V_\text{inf}(\varphi) - m_{3/2}^2 \frac{f_1^2(\varphi) + f_2^2(\varphi)}{M_S^2(\varphi)}\,.
\end{align}
The backreaction term is always negative and is generically problematic for inflation even if an appropriate uplift is taken into account.\footnote{Notice that, in principle, the backreaction term can be made to vanish by making the mass of the stabilizer field very large. As will become clear in the examples we discuss, this is unrealistic in most scenarios due to the origin of the mass term of the stabilizer field.} This has been studied in detail for the specific case of chaotic inflation with a stabilizer field, i.e., the model first proposed in \cite{Kawasaki:2000yn}, in \cite{Buchmuller:2014pla}. As we discuss in this paper, the results found in \cite{Buchmuller:2014pla} hold more generally. An exception seems to be models of natural inflation, which are somewhat protected from severe backreactions by the periodicity of the inflaton potential. We discuss these models as well as other examples in Sec.~3.

%
%%
%%%
%%%%
%%%%%%%%%%%%%%%%%%%%%%%%%%%%%%%%%%%%%%%%%%%%%%%
%%%%%%%%%%%%%%%%%%%%%%%%%%%%%%%%%%%%%%%%%%%%%%%
\subsection{Backreaction of heavy moduli}

Supergravity models without stabilizer fields in the presence of geometric moduli have been recently studied in \cite{Buchmuller:2015oma} with particular focus on chaotic inflation with a quadratic potential. The authors have demonstrated that there are non-decoupling effects induced by a backreaction of the moduli and by supersymmetry breaking which become more severe as the moduli become heavier. Thus, their influence on the inflationary dynamics are never negligible and are potentially destructive.  We wish to stress here that the results found in \cite{Buchmuller:2015oma} hold more generally and for a much larger class of inflation models. This has implications for many string-derived and string-inspired models in the recent literature.

Along the lines of \cite{Buchmuller:2015oma} we can start our analysis from a quite general ansatz of the form\footnote{One important class of models which is not captured by this ansatz is K\"ahler moduli inflation in its various forms, cf.~\cite{Baumann:2014nda} for a review and a list of references. In many of those models, like the ones of \cite{Conlon:2005jm,Cicoli:2008gp,Cicoli:2011ct,Blumenhagen:2012ue}, the inflaton potential has a plateau as in some of our examples. Most of them take various backreactions from heavy fields into account.}
\begin{align}\label{eq:Kahleransatz2}
K &= K_0 (T_\alpha,{\overline T}_\alpha) + K_1 (\Phi + {\overline \Phi}, X, \bar X, T_\alpha,\overline T_\alpha) \,, \\
W &= W_{\rm inf} (\Phi) + W_1 (X,T_\alpha) \,. \label{g4}
\end{align}
Once more the inflaton field $\varphi$ is protected by a shift symmetry, $X$ is responsible for an uplift of the true vacuum and $T_\alpha$ denotes all moduli fields. Thus, $K_0$ usually has a no-scale symmetry up to perturbative corrections from string theory. The mixing between the moduli and the inflaton is assumed to be purely gravitational.

The backreaction of moduli on the inflaton potential can be worked out by expanding the scalar potential in
\begin{equation}
V = V_0 + V_1 + V_2 + \dots \,,  \label{g5} 
\end{equation}
where $V_0$ denotes the moduli potential at the end of inflation, ${V_1 \sim \mathcal O(W_{\rm inf})}$, and ${V_2 \sim  \mathcal O(W_{\rm inf}^2)}$. This means we consider the case that the moduli fields are much heavier than the inflaton and treat inflation as a perturbation of the modulus potential.

According to the analysis in Sec.~2 of \cite{Buchmuller:2015oma} we can treat the potential as a perturbative expansion in the displacement of the moduli fields, since during inflation $T_\alpha = T_{\alpha,0} + \delta T_\alpha(\varphi)$. Here $T_{\alpha,0}$ denotes the vacuum expectation value of $T_\alpha$ after inflation has ended. Similar to the models with heavy stabilizer fields we can integrate out the moduli to find the effective inflaton potential, along the lines of \cite{Buchmuller:2015oma}. The important difference in our case is that we expand in a general superpotential $W_\text{inf}$ instead of in $\varphi$. We can write the result as 
\begin{align}
V (\varphi) = \Lambda_0^4 + V_1 (T_{\alpha,0}, {\overline T}_{\alpha,0},\varphi ) + V_2 (T_{\alpha,0}, {\overline T}_{\alpha,0},\varphi ) - \frac{1}{2} \frac{\partial V_1}{\partial \rho_\alpha}  M_{\alpha \beta}^{-2}
\frac{\partial V_1}{\partial \rho_\beta} + \dots \,, \label{g9} 
\end{align}
with $\rho_\alpha = (T_\alpha, \overline T_\alpha)$. We have summarized the effect of the uplift sector in a constant $\Lambda_0$ which cancels the cosmological constant in the true vacuum after inflation. The second term on the right-hand side arises through supersymmetry breaking and is the analog of soft terms in phenomenological models like the MSSM and the soft inflaton mass term in \cite{Buchmuller:2015oma}. Its explicit form in this general setup is irrelevant, and we discuss examples in Sec.~4. The last term in Eq.~\eqref{g9} is the effect of the moduli backreaction. It is suppressed by the squared inverse of the modulus mass matrix. Nevertheless it contains additional non-decoupling effects which survive in the limit of large moduli masses, if the latter contribute to supersymmetry breaking.

We can provide more explicit expressions in a class of models where the kinetic term of the inflaton is simple,
\begin{align}
K &= K_0 (T_{\alpha}, {\overline T}_{\bar \alpha} ) + \frac{1}{2} K_1 (T_{\alpha}, {\overline T}_{\bar \alpha}) (\Phi + {\overline \Phi})^2 \,, \\
 W &= W_{\rm \inf} (\Phi) +  W_\text{mod}  (T_{\alpha}) \,, \label{g10}
\end{align}
neglecting the uplift sector involving the field $X$ for the moment.\footnote{In \cite{Buchmuller:2015oma} it was shown that including the supersymmetry breaking sector in the analysis does not change the final result, provided the sgoldstino is heavy enough and has a small vacuum expectation value.} Integrating out the displacement of the moduli and expanding in the inflaton superpotential yields\footnote{In\cite{Buchmuller:2015oma} the supersymmetric mass of order $\mathcal O(W_{\rm inf}^2)$ was included in $V_1$, whereas according to our present expansion it should have been included in $V_2$. Consequently, a sub-leading term of order $\mathcal O(W_{\rm inf}^3)$ was kept in the final expression, Eq.~(2.16) of that paper.} for the terms in Eq.~\eqref{g5}
\begin{align}
V_1 &= e^{K_0}  \left\{ \left( K_0^{\alpha \bar \beta} K_{0, \alpha } \overline{D}_{\bar \beta} \overline{W}_\text{mod} - 3  \overline{W}_\text{mod} \right) W_{\rm \inf} (\Phi) + {\rm h.c.} \right\} \,, \\
V_2  &=  e^{K_0}  \left\{ \left( K_0^{\alpha \bar \beta} K_{0, \alpha } K_{0, {\bar \beta} } - 3\right) |W_{\rm \inf} (\Phi)|^2 + K_1^{-1} |\partial_\Phi W_{\rm \inf} (\Phi)|^2   \right\}  \,, \label{g15}
\end{align}
where $D_{\alpha} W_\text{mod} = W_{\text{mod}, \alpha} + K_{0, \alpha } W_\text{mod}$. The corresponding result for Eq.~\eqref{g9} is still a very complicated expression. Using the explicit form of $\partial V_1/\partial T_\alpha$ we can simplify the result by assuming that the influence of the field which breaks supersymmetry dominantly is weak. This is the case when the supersymmetric mass of the chiral fields in the theory is much greater than the gravitino mass, or when the supersymmetry breaking scale is large but the supersymmetry-breaking sector decouples from moduli stabilization by some mechanism. The latter is the case, for example, in KKLT moduli stabilization with a Polonyi uplift. This is the case we mostly consider in the example models of Sec.~4. More details on this approximation and the simplification of the effective potential can be found in the appendix of \cite{Buchmuller:2015oma}. The result of this computation reads
\begin{align}
&V (\varphi) = \Lambda_0^4 + e^{K_0} \bigg\{ K_1^{-1} |\partial_\Phi W_{\rm \inf} (\Phi)|^2 - 3 \left|W_{\rm \inf} (\Phi) \right|^2  \nonumber \\ 
& \hspace{2.5cm}+ \left[ \left (K_0^{\alpha \bar \beta} K_{0,\alpha } \overline{D}_{\bar \beta} \overline{W}_\text{mod} - 3 \overline{W}_\text{mod} \right)  W_{\rm \inf} (\Phi) + {\rm h.c.} \right] \bigg\}   \nonumber  \\
& \hspace{1cm}+ e^{\frac32 K_0} \biggl( K_{\delta} \left(m_F^{-1}\right)^{\beta \delta} 
\bigg\{ - \Big[ K_0^{\epsilon \bar \epsilon} (K_{\beta \epsilon} + K_{\beta} K_{\epsilon} - \Gamma_{\beta \epsilon}^{\gamma} K_{\gamma}) \overline{D}_{\bar \epsilon} \overline{W}_\text{mod} \nonumber \\
& \hspace{4.3cm}- 3 K_{\beta} {\overline W}_\text{mod}  \Big]  W_{\rm \inf}^2(\Phi) 
+ 2 D_{\beta} W_\text{mod}  |W_{\rm \inf} (\Phi)|^2 \bigg\} + \text{h.c.} \biggr)  \nonumber\\
& \hspace{1cm} + \mathcal O \left( \frac{H^2}{m_T^2} \right) \,, \label{g17} 
\end{align}
with $\Gamma^\alpha_{\beta \gamma} = G^{\alpha \bar \alpha}Ê\partial_\beta G_{\gamma \bar \alpha}$ and $G = K + \ln|W|^2$. Moreover, $m_F^{-1}$ denotes the inverse of the fermion mass matrix,
\begin{align}
(m_F)_{\alpha \beta} = e^{G/2} \left(  G_{\alpha \beta} - \Gamma^\gamma_{\alpha \beta} G_\gamma + \frac13 G_\alpha G_\beta \right)\,,
\end{align}
which determines the supersymmetric contribution to the scalar masses. Again $\Lambda_0$ summarizes the effect of an additional uplift sector. Notice that the term in the second line, proportional to $W_\text{inf}$, is the soft term discussed before. It is independent of any approximations and assumptions made about the scale of supersymmetry breaking. Moreover, in Eq.~\eqref{g17} $\Phi = i (2K_1)^{-1/2} \varphi$ so that $\varphi$ is the canonically normalized inflaton field, and the real part of $\Phi$ is stabilized at the origin by its soft mass term.

In Sec.~4 we provide various examples of the corrections to the inflaton potential in single-field inflation models, by working out Eq.~\eqref{g9} and  Eq.~\eqref{g17} explicitly, whenever applicable. Before we proceed with examples of models with stabilizer fields, let us comment briefly on possible higher-order correction to the inflaton potential.

\subsection{Comments on higher-order corrections}

In any UV complete theory which produces the low-energy effective supergravity setups we consider, one may worry about the effect of additional Planck-suppressed corrections and their effect on the dynamics of inflation. Specifically, our two classes of models in Eqs.~\eqref{eq:Kahleransatz1} and \eqref{eq:Kahleransatz2} assume that the inflaton field is protected by a shift symmetry. This happens generically in various type II string compactifications where the shift symmetry is a remnant of a ten-dimensional gauge symmetry. In heterotic string compactifications it may arise as a remnant of $SL(2,\mathbb Z)$ symmetries of the tori of the compact manifold. In both cases the shift symmetry is exact at all orders in perturbation theory, i.e., we expect perturbative corrections to $K$ in Eqs.~\eqref{eq:Kahleransatz1} and \eqref{eq:Kahleransatz2} to respect the shift symmetry.\footnote{For a recent discussion of this point we refer the reader to \cite{Bielleman:2015ina}.}

Non-perturbative corrections, usually proportional to $e^{-\alpha \Phi}$, may break the shift symmetry to a discrete subgroup, as it happens in our examples of natural inflation. In those cases we assume that additional instantonic contributions to the low-energy effective action are smaller than those which generate the inflaton or moduli potentials.

Furthermore, as demonstrated in \cite{Kaloper:2008fb,Kaloper:2011jz,Kaloper:2014zba}, higher-derivative interactions which are not captured by the setups of Eqs.~\eqref{eq:Kahleransatz1} and \eqref{eq:Kahleransatz2} are expected to scale with powers of the original potential, $V_0^n/M_\text{P}^{4(n-1)}$, and are thus generically suppressed compared to the corrections we discuss in this paper.

%
%%
%%%
%%%%
%%%%%%%%%%%%%%%%%%%%%%%%%%%%%%%%%%%%%%%%%%%%%%%
%%%%%%%%%%%%%%%%%%%%%%%%%%%%%%%%%%%%%%%%%%%%%%%
%%%%%%%%%%%%%%				    %%%%%%%%%%%%%%%%%%%%%%%
%%%%%%%%%%%%%%         Section 3        %%%%%%%%%%%%%%%%%%%%%%%
%%%%%%%%%%%%%%				    %%%%%%%%%%%%%%%%%%%%%%%
%%%%%%%%%%%%%%%%%%%%%%%%%%%%%%%%%%%%%%%%%%%%%%%
%%%%%%%%%%%%%%%%%%%%%%%%%%%%%%%%%%%%%%%%%%%%%%%

\section{Models with stabilizer fields}

Stabilizer fields have first been used in context of inflation in \cite{Kawasaki:2000yn}.\footnote{Note also the enlightening discussion in \cite{Roest:2013aoa} on possible K\"ahler potentials in models with stabilizer fields. A microscopic embedding of models with stabilizer fields has been recently proposed in \cite{Dudas:2014pva}.} Unsurprisingly, they stabilize the inflationary trajectory and help to achieve Minkowski or de Sitter vacua after inflation. This is because they usually enter the superpotential linearly and are stabilized at the origin, causing $\langle W \rangle = 0$ in the vacuum so that $\langle V \rangle \geq 0$. As pointed out in Sec.~2, however, their interplay with supersymmetry breaking induces correction terms in the inflaton potential. In the following we analyze these in plateau inflation models and natural inflation models, respectively.

%
%%
%%%
%%%%
%%%%%%%%%%%%%%%%%%%%%%%%%%%%%%%%%%%%%%%%%%%%%%%
\subsection{Starobinsky-like models with stabilizer fields}
\label{sec:Stabilizer}

In the original proposal of \cite{Starobinsky:1980te} inflation is driven by the vacuum energy of a scalar field whose potential is generated by a non-minimal coupling to gravity. It reads
\begin{align}
V = \Lambda^4\left(1-e^{-\sqrt{\frac23} \varphi}\right)^2\,,
\end{align} 
where $\Lambda$ denotes the energy density during inflation and $\varphi$ is the real inflaton field. The potential is exponentially flat at large inflaton field values and produces observables in accordance with the most recent CMB data. Much effort has been devoted recently to embedding this scenario in four-dimensional $\mathcal N =1$ supergravity. One important class of plateau-like models in supergravity are those including a stabilizer field. While stability of all directions in field space is usually not an issue in those cases, they are generically constrained once supersymmetry breaking is taken into account. In what follows we illustrate this by means of two examples.

%
%%
%%%
%%%%
%%%%%%%%%%%%%%%%%%%%%%%%%%%%%%%%%%%%%%%%%%%%%%%
\subsubsection{The modified Cecotti model}

One of the first implementations of a plateau-like model in no-scale supergravity was proposed in \cite{Cecotti:1987sa} and further developed in \cite{Kallosh:2013lkr}. The model is comprised of a chiral superfield $\Phi$ containing the inflaton field and a stabilizer field $S$ which obey the following action,
\begin{align}\label{eq:CecK}
K &= -3 \log \left( \Phi + \overline \Phi  - |S|^2 + \frac{\xi}{3} |S|^4 \right)\,,\\
W &= M S (\Phi-1)\,.
\end{align}
$M$ is a mass scale which determines the energy scale of inflation. The quartic term for the stabilizer field was introduced in \cite{Kallosh:2013lkr} and is necessary for $S$ to decouple sufficiently during inflation. Notice that a kinetic term of $S$ outside the logarithm would not give a plateau model. During inflation $S$ is stabilized at the origin while slow-roll proceeds along the direction of the canonically normalized inflaton field $\varphi$, defined by $\Phi = e^{\sqrt{\frac23} \varphi} + i a$. The field $a$ is stabilized at the origin with a Hubble-scale soft mass and decouples from inflation. The scalar potential on the inflationary trajectory thus becomes
\begin{align}\label{eq:CecV}
V = \frac{M^2}{12}\left( 1- e^{-\sqrt{\frac23} \varphi} \right)^2\,.
\end{align}
This is the original Starobinsky potential with a plateau at large field values, $\varphi \gtrsim 5$. However, in the vacuum after inflation, corresponding to $\langle \varphi \rangle = 0$ or $\langle \Phi \rangle = 1$, the potential vanishes and supersymmetry is unbroken. Since control over the scale of supersymmetry breaking is desirable from a phenomenological perspective, in the following we study the effects of F-term supersymmetry breaking on the dynamics of inflation.

%
%%
%%%
\paragraph{Supersymmetry breaking and backreaction of the stabilizer field}
$\,$\\
As in other inflation models involving stabilizer fields we expect that $S$ backreacts on the inflaton potential if the supersymmetry breaking scale is large. For chaotic inflation with a stabilizer field this was shown in \cite{Buchmuller:2014pla}. The simplest and least constraining way to incorporate supersymmetry breaking is via a Polonyi field, i.e.,
\begin{align}\label{eq:CecK}
K &= -3 \log \left( \Phi + \overline \Phi  - |S|^2 + \frac{\xi}{3} |S|^4 \right) + k(|X|)\,,\\
W &= M S (\Phi-1) + fX + W_0\,.
\end{align}
We assume that the function $k$ can be chosen such that $X$ is stabilized near the origin of field space and decouples from inflation. This happens, for instance, through one-loop interactions with other heavy fields, cf.~\cite{Grisaru:1996ve} for more details. Thereby the dynamics of the Polonyi field can be completely decoupled from inflation.\footnote{Alternatively, such a decoupling can be achieved by imposing a nilpotency condition on $X$, cf.~the discussions in \cite{Antoniadis:2014oya,Ferrara:2014kva,Kallosh:2014via,Dall'Agata:2014oka}. This may work similarly for the stabilizer field $S$ by imposing $X S = 0$.} Notice that a kinetic term of $X$ inside the logarithm would not work since it would spoil the plateau.

To evaluate the effects of the supersymmetry breaking sector we can write the stabilizer field as $S = \chi + i \beta$ and expand the potential around the origin in $\chi$ and $\beta$. The result reads
\begin{align}
V = V_0 + \frac{e^{-3\sqrt{\frac23}\varphi}}{8} f^2 - \frac{e^{-2\sqrt{\frac23}\varphi}}{2} M W_0 \chi +  m_1^2 \chi^2 + m_2^2 \beta^2\,,
\end{align}
where $V_0$ denotes the original potential in Eq.~\eqref{eq:CecV} and $m_1$ and $m_2$ denote inflaton-dependent mass terms for $\chi$ and $\beta$, respectively. Apparently $\beta$ remains stabilized at the origin while the minimum of $\chi$ is displaced due to the linear term proportional to $W_0$. This term is the explicit form of the first term in Eq.~\eqref{g2}, with $m_{3/2} \sim W_0$. Notice that, while cosmological constant cancellation in the true vacuum implies a relation between $f$ and $W_0$, there can be no Minkowski solution with $\langle \varphi \rangle = \langle \chi \rangle = 0$ and broken supersymmetry. Using the explicit form of $m_1$ we can solve for the displacement $\delta \chi(\varphi)$ to determine the backreaction. We find 
\begin{align}\label{eq:CecShift}
\delta \chi \approx \frac{9 W_0}{4 \xi M} e^{-2\sqrt{\frac23}\varphi}\,.
\end{align}
Notice that this shift decreases as $\varphi$ increases, and thus vanishes in the plateau regime of the potential. Nevertheless, it affects the potential for small inflaton field values and may interfere with inflation. Inserting Eq.~\eqref{eq:CecShift} into $V$ yields the effective potential for the inflaton,
\begin{align}
V = V_0 +  \frac{f^2}{8} e^{-3\sqrt{\frac23}\varphi}- \frac{9 W_0^2}{8 \xi} e^{-4\sqrt{\frac23}\varphi}\,,
\end{align}
at leading order in $\delta \chi$. The true vacuum of the theory can in principle be found by solving $\partial_\varphi V = V = 0$. Since the correction terms are suppressed for large inflaton field values it seems that there is no constraint on this theory at all. However, we must still require that $\delta \chi \ll 1$ at all times, i.e., for all values of $\varphi$ including $\varphi = 0$. This is to guarantee that the expansion converges and to ensure that perturbative corrections to the K\"ahler potential are under control.\footnote{Moreover, the K\"ahler metric exhibits a singularity at $|S|^2 \sim 2 \text{Re}\,\Phi \sim 2$.} For $\xi \sim \mathcal O(1)$, which is realistic when the quartic term in the K\"ahler potential arises from interactions with heavy fields, this implies
\begin{align}
W_0 \ll M\,.
\end{align}
Since $M \sim 10^{-5}$ is fixed by observations, the allowed gravitino mass is bounded from above by
\begin{align}\label{eq:CecBound}
m_{3/2} < 10^{13}\,\text{GeV} \sim H\,.
\end{align}
This bound is very similar to the one found in chaotic inflation in the analysis of \cite{Buchmuller:2014pla}.

%
%%
%%%
\paragraph{Comments on moduli stabilization}
$\,$\\
The result found above has implications for possible string theory embeddings of this model. In many string compactifications supersymmetry is generically broken by fluxes close to the string scale or the GUT scale, or alternatively by the auxiliary fields of K\"ahler moduli. The latter case, as realized in KKLT or the Large Volume Scenario, usually requires at least ${m_{3/2} > H}$ for moduli stability, cf.~the discussion in Sec.~\ref{sec:ENO}. Although in the present model $\Phi$ enters the K\"ahler potential like a K\"ahler modulus parameterizing the entire compactification volume, its string theory interpretation is unclear since it appears perturbatively in the superpotential.\footnote{Under certain circumstances such superpotentials for moduli may be generated by closed-string fluxes.} However, if a suitable model with the correct couplings was to be found in string theory, all other remnant moduli would have to be stabilized by some mechanism. The result in \eqref{eq:CecBound} implies that this mechanism can not be KKLT or one of its variants. Instead, it seems that one is forced to strong moduli stabilization, as discussed in Sec.~\ref{sec:ENOstrong}, where the modulus mass and the gravitino mass are unrelated.

%
%%
%%%
%%%%
%%%%%%%%%%%%%%%%%%%%%%%%%%%%%%%%%%%%%%%%%%%%%%%

\subsubsection{A model with an analytic K\"ahler metric}

As our last example of this class, let us turn to a plateau model with stabilizer field which does not exhibit a singularity in the K\"ahler metric. We choose
\begin{align}
K &= -2 \log\left( \Phi + \overline \Phi \right) + |S|^2 - \xi|S|^4\,,\\
W &= M S (\Phi - \Phi^2)\,.
\end{align}
Again a kinetic term of $S$ inside the logarithm would not preserve the plateau. Analogous to the previous example the potential on the inflationary trajectory with $S = 0$ and $\Phi = e^{-\varphi}$ reads\footnote{Again the imaginary part of $\Phi$ is fixed at the origin with a Hubble-scale soft mass and decouples from inflation.}
\begin{align}\label{eq:EmilianV}
V = \frac{M^2}{4}\left(1-e^{-\varphi}\right)^2\,.
\end{align}
Again, in the vacuum supersymmetry is unbroken so that an analysis of the effects of a supersymmetry-breaking sector becomes necessary.

%
%%
%%%
\paragraph{Supersymmetry breaking and backreaction of the stabilizer field}
$\,$\\
Once more we study the interaction of inflation and supersymmetry breaking using the ansatz
\begin{align}
K &= -2 \log\left( \Phi + \overline \Phi \right) + |S|^2 - \xi|S|^4\ + k(|X|)\,,\\
W &= M S (\Phi - \Phi^2) + fX + W_0\,.
\end{align}
Once more, the kinetic term of $X$ must not be inside the logarithm to obtain a plateau in the potential.
Cosmological constant cancellation in the true vacuum with $\varphi = 0$ requires $f = W_0$. Similar to the previous example we can write $S = \chi + i \beta$ and observe that a linear term in $\chi$ induces a backreaction in the inflaton potential. Analogous to Eq.~\eqref{eq:CecShift} we find 
\begin{align}
\delta \chi \approx -\frac{2 M W_0 \left(2 - e^{\varphi} \right)}{W_0^2\, e^{2 \varphi}+ 2 M^2 \, e^{-2 \varphi} + 4 M^2 \xi \left(1-e^{- \varphi}\right)^2}\,.
\end{align}
In contrast to the Cecotti model, in this case the backreaction becomes stronger with increasing value of $\varphi$. Thus, we expect the plateau to be affected by the backreaction. Using this result the effective inflaton potential takes the form
\begin{align}\label{eq:back1}
V = V_0 - \frac{M^2 W_0^2 \left(2 - e^{ \varphi} \right)^2}{W_0^2\, e^{2 \varphi}+ 2 M^2 \, e^{-2 \varphi} + 4 M^2 \xi \left(1-e^{- \varphi}\right)^2}\,,
\end{align}
where $V_0$ denotes the original potential in Eq.~\eqref{eq:EmilianV}. Note that, in line with the discussion in Sec.~\ref{sec:GenStab}, the denominator of the correction term is the squared mass term of $S$. As mentioned before, it can not be made arbitrarily large to suppress the backreaction. $M$ is fixed by observations and $\xi$ is bounded by consistency of the effective field theory, cf.~the more detailed treatment in \cite{Buchmuller:2014pla}.

The negative correction term in Eq.~\eqref{eq:back1} dominates for large inflaton field values. Depending on the magnitude of $W_0$ it may dominate even for $\varphi < \varphi_\star \sim 6$, which is the starting point of the last 60 $e$-folds of slow roll inflation. In particular, in the limit $\varphi \to \infty$ the correction term becomes $-M^2$. For $V_0$ to be larger than the correction term it must be 
\begin{align}
W_0 < M \sqrt{\xi}\,e^{- \varphi}\,.
\end{align}
With $\xi \sim \mathcal O(1)$ and $\varphi = \varphi_\star \sim 6$ this becomes $W_0 < 10^{-7}$, a quite conservative bound. In fact, the cosmological observables are affected for even smaller values of $W_0$.
%%%
\begin{figure}[t]
\centering
\includegraphics[width=1\textwidth]{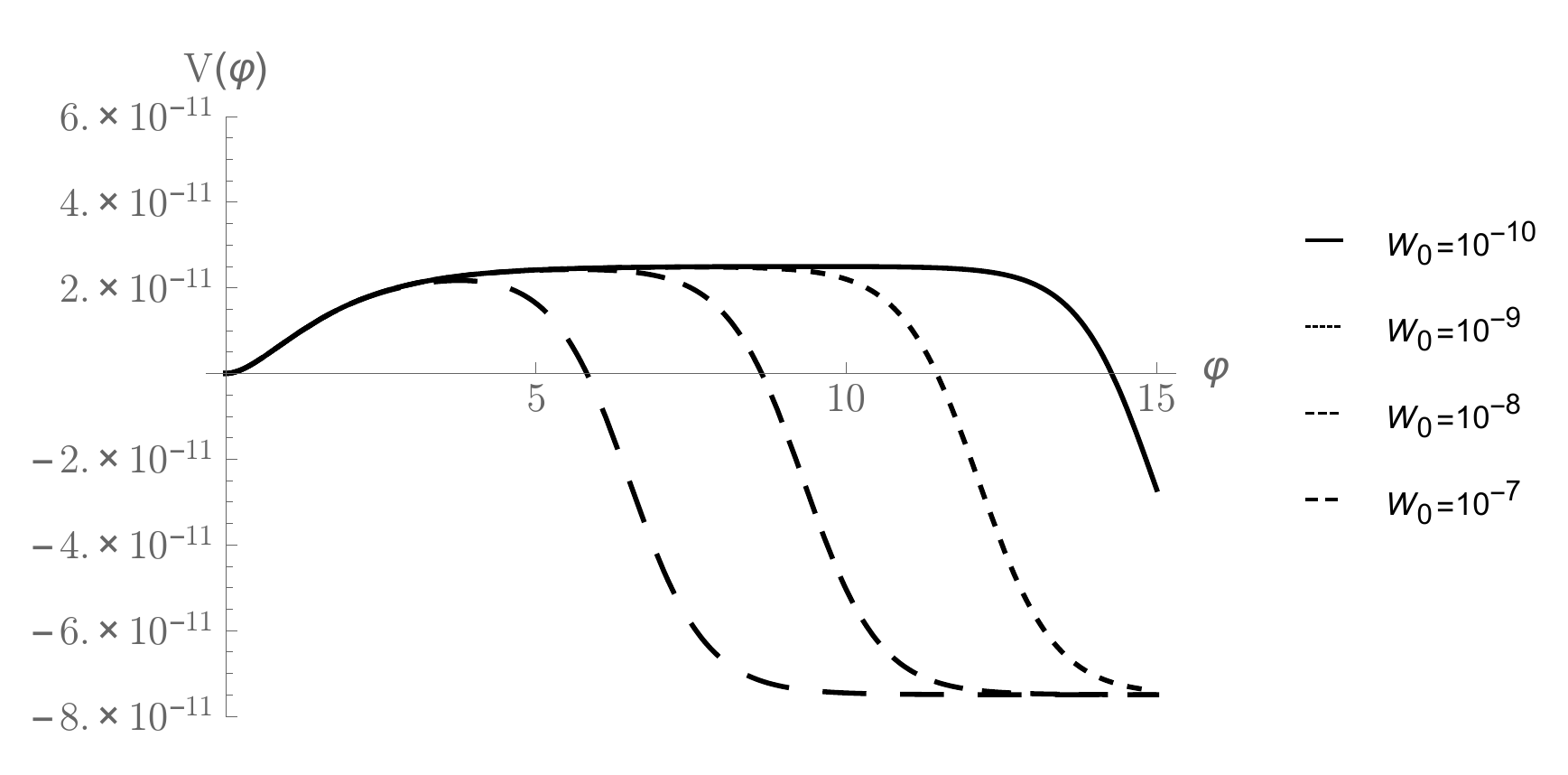} 
\caption{Effective inflaton potential for $M = 10^{-5}$, $\xi = 1$, and $W_0$ ranging from $10^{-10}$ to $10^{-7}$. Evidently, the inflationary plateau is always destroyed for large field values, i.e., $V \to -\frac34 M^2$ as $\varphi \to \infty$.
\label{fig:effpot}}
\end{figure}
%%%
We have illustrated the effective potential with $M = 10^{-5}$ and $\xi = 1$ for several values of $W_0$ in Fig.~\ref{fig:effpot}. Apparently, for $W_0 \gtrsim 10^{-9}$ the plateau becomes quite narrow so that the initial conditions must be chosen very carefully. If $W_0 \gtrsim 10^{-8}$ the hilltop is already too close to $\varphi_\star$ for 60 $e$-folds of inflation to be possible in agreement with CMB observations. Thus, it seems that the upper bound
\begin{align}
m_{3/2} < 10^{10}\,\text{GeV}\,,
\end{align}
must be satisfied for inflation to be viable in this setup. Since the backreaction affects the plateau regime of the potential, this bound is even more severe than the one found in the previous example. Even for substantially smaller gravitino masses the plateau is always lost at very large field values, foiling one of the prime virtues of plateau-like inflation. Moreover, this result implies that the comments about moduli stabilization made in the previous example apply in this case as well. 

%
%%
%%%
%%%%
%%%%%%%%%%%%%%%%%%%%%%%%%%%%%%%%%%%%%%%%%%%%%%%
%%%%%%%%%%%%%%%%%%%%%%%%%%%%%%%%%%%%%%%%%%%%%%%
\subsection{Natural inflation models with stabilizer fields}
\label{sec:NatStabilizer}

In natural inflation the expansion of the early universe is driven by an axionic field with a discrete shift symmetry. Thus, a periodic potential for the axion arises, i.e.,
\begin{align}
V = \Lambda^4\left[1- \cos\left(\frac{\varphi}{f}\right)\right]\,.
\end{align}
For a sufficiently large axion decay constant $f$ the predicted observables of this model are in agreement with the most recent CMB data. Achieving super-Planckian decay constants, $f \gtrsim 1$, is a subtle issue. We postpone this discussion and first study a minimal toy model of natural inflation involving a stabilizer field. As in the case of plateau inflation the backreaction of the stabilizer field becomes important when the scale of supersymmetry breaking becomes large. Afterwards we comment on possibilities to enhance the effective axion decay constant. 

%
%%
%%%
%%%%
%%%%%%%%%%%%%%%%%%%%%%%%%%%%%%%%%%%%%%%%%%%%%%%
\subsubsection{Natural inflation and large axion decay constants}
\label{sec:NatStab1}

A simple way to realize natural inflation in supergravity has been developed in \cite{Kallosh:2014vja}. The corresponding Lagrangian for two chiral multiplets $\Phi$ and $S$ is defined by 
\begin{align}\label{eq:NatLag}
K &= \frac12(\Phi + \overline \Phi) + |S|^2 - \xi |S|^4 \,,\\ 
W &= M^2 S \left(1- e^{-a \Phi}\right)\,.
\end{align}
In string theory a superpotential like this may arise from Yukawa couplings of matter fields and couplings to world-sheet or space-time instantons whose action is determined by $\Phi$. As before, during inflation the stabilizer field is fixed at the origin and its auxiliary field sources the inflaton potential,
\begin{align}\label{eq:NatPot1}
V = 2 M^4 \left[1-\cos\left( \frac{a \varphi}{\sqrt 2} \right)\right]\,.
\end{align}
Here $\varphi = \sqrt 2 \,\text{Im}\, \Phi$ denotes the canonically normalized inflaton field. The real part of $\Phi$ is stabilized close to the origin with a Hubble-scale mass.

However, once supersymmetry breaking is taken into account the potential changes. Let us consider, once more, a coupling to a Polonyi field $X$ of the type
\begin{align}
K &= \frac12(\Phi + \overline \Phi) + |S|^2 - \xi |S|^4 + k(|X|)\,,\\ 
W &= M^2 S \left(1- e^{-a \Phi}\right) + f X + W_0\,.
\end{align}
The additional F-term again induces inflaton-dependent linear terms for both $\text{Re}\, S$ and $\text{Im}\,S$ so that the stabilizer backreacts on the inflationary trajectory. As before we integrate out $S$ and find for the effective inflaton potential
\begin{align}\label{eq:NatEffPot1}
V = f^2- 3 W_0^2 + V_0 - \frac{4 W_0^2 V_0}{f^2 - 2W_0^2 + a^2 M^4 + 4 \xi V_0}\,,
\end{align}
with $V_0$ given by Eq.~\eqref{eq:NatPot1}. There are two interesting cases to be considered here. If $W_0$ -- and hence the scale of supersymmetry breaking -- is small, $W_0 \ll M^2 \sim H$, cancellation of the cosmological constant in the vacuum is enforced by $f = \sqrt 3 W_0$ and the last term in \eqref{eq:NatEffPot1} is subdominant. In this case natural inflation may proceed unperturbed. If, on the other hand, $W_0$ is large compared to $H$ the relation between $f$ and $W_0$ slightly changes and the correction term in \eqref{eq:NatEffPot1} becomes approximately $-4 V_0$. The resulting effective potential is again a cosine function with the same frequency as before.\footnote{Its amplitude differs by a factor of three, which means that the inflationary energy density differs by a factor of $3^{1/4}$. This can be compensated by a redefinition of $M$.} With an appropriate uplift, by choice of $f$, inflation may proceed driven by the effective potential
\begin{align}
V = 6 M^4 \left[1 + \cos\left( \frac{a \varphi}{\sqrt 2} \right)\right]\,.
\end{align}
Hence, there is an important difference to Starobinsky-like inflation or chaotic inflation. Due to the periodicity of the scalar potential, also the correction induced by the backreaction of the stabilizer field is periodic and may be brought back to the same form as the original potential. We remark that there is an intermediate regime $W_0 \sim H$, in which the shape of the potential is changed significantly so that inflation may not be possible.

A point we have not addressed so far is the issue of super-Planckian axion decay constants. With the above scalar potential, agreement with observations is only possible if $a \ll1$. However, string theory suggests that $aÊ\gtrsim 1$ \cite{Banks:2003sx,Svrcek:2006yi}. The model defined by Eqs.~\eqref{eq:NatLag} can be extended in a simple way to achieve an effectively super-Planckian axion decay constant. As noted in \cite{Kallosh:2014vja}, including a second non-perturbative term involving $\Phi$, i.e.,
\begin{align}\label{eq:NatSupPot3}
W = M^2 S (A + B e^{-a \Phi} + C e^{-b \Phi})\,,
\end{align}
leads to an inflationary potential of the form 
\begin{align}\label{eq:NatPot2}
V = M^4 \left[ A^2+B^2+C^2 + 2 A B \cos \left( \frac{a \varphi}{\sqrt 2}\right) + 2 A C \cos \left( \frac{b \varphi}{\sqrt 2}\right) + 2 B C \cos \left( \frac{(a-b) \varphi}{\sqrt 2}\right)\right]\,.
\end{align}
If the constant coefficients are chosen such that $BC < 0$ and $A \ll B,C$ the last term in Eq.~\eqref{eq:NatPot2} may drive inflation. Then, even if $a,b \gtrsim 1$ it is possible that $a-b \ll 1$ so that the inflaton has an effectively super-Planckian decay constant. When this theory is coupled to supersymmetry breaking the effect of the backreaction of the stabilizer is precisely the same as in the previous case. Thus, natural inflation is possible in both regimes $W_0 \ll H$ and $W_0 \gg H$.

Another option to enhance the effective decay constant on the inflaton trajectory is by aligning two axion fields in an appropriate way \cite{Kim:2004rp}. Models of aligned natural inflation have recently been implemented in string-motivated supergravity using stabilizer fields in \cite{Kappl:2015pxa,Ruehle:2015afa}. Also in those cases the backreaction of the stabilizer becomes sizeable when $W_0 \lesssim H$. But again, due to the periodicity of the potential, inflation is still possible for large values of $W_0$ if an appropriate uplift is taken into account. 

%
%%
%%%
%%%%
%%%%%%%%%%%%%%%%%%%%%%%%%%%%%%%%%%%%%%%%%%%%%%%
\subsubsection{Natural inflation and monodromy}

Yet another possibility to reconcile natural inflation with CMB data is by introducing a monodromy for the shift-symmetric axion field \cite{Silverstein:2008sg,McAllister:2008hb}. Regardless of the string theory embedding of such models, in the low-energy effective supergravity potential such monodromies usually imply lifting the periodic potential with a monomial function. One possible setup involving a stabilizer field, based on the previous example, can be defined by 
\begin{align}\label{eq:NatLag2}
K &= \frac12(\Phi + \overline \Phi) + |S|^2 - \xi |S|^4 \,,\\ 
W &= M^2 S \left(\Phi- A e^{-a \Phi}\right)\,,
\end{align}
where $A$ is a constant parameter. The inflaton potential becomes 
\begin{align}\label{eq:NatPot3}
V = M^4 \left[A^2+ \frac12 \varphi^2 + \sqrt 2 A \varphi \sinÊ\left( \frac{a \varphi}{\sqrt 2} \right)  \right]\,.
\end{align}
The result is a quadratic potential with sinusoidal modulations. Coupling this model to a Polonyi field works the same way as in the previous examples. The backreaction of $S$ again becomes important when $W_0 \lesssim M^2 \sim H$. The corrected inflaton potential is identical to Eq.~\eqref{eq:NatEffPot1}, but with $V_0$ as in Eq.~\eqref{eq:NatPot3}. This time, if $W_0 \gg M^2$ inflation is impossible since the negative correction term does not have the same form as the original potential. This is because the periodicity of the potential has been lifted by the quadratic term. Thus, the same conclusions as in chaotic inflation with a quadratic potential apply, cf.~the more thorough analysis in \cite{Buchmuller:2014pla}. In this case, there is indeed an upper bound on the gravitino mass,
\begin{align}
m_{3/2} < H\,.
\end{align}

%
%%
%%%
%%%%
%%%%%%%%%%%%%%%%%%%%%%%%%%%%%%%%%%%%%%%%%%%%%%%
%%%%%%%%%%%%%%%%%%%%%%%%%%%%%%%%%%%%%%%%%%%%%%%
%%%%%%%%%%%%%%				    %%%%%%%%%%%%%%%%%%%%%%%
%%%%%%%%%%%%%%         Section 4        %%%%%%%%%%%%%%%%%%%%%%%
%%%%%%%%%%%%%%				    %%%%%%%%%%%%%%%%%%%%%%%
%%%%%%%%%%%%%%%%%%%%%%%%%%%%%%%%%%%%%%%%%%%%%%%
%%%%%%%%%%%%%%%%%%%%%%%%%%%%%%%%%%%%%%%%%%%%%%%
\section{Models without stabilizer fields}

Backreactions of heavy fields are not only important concerning the stabilizer field. There exist numerous setups without stabilizer fields which reproduce natural inflation or plateau-like inflation in some limit. In the following we would like to discuss a few representative examples. In particular, we wish to stress the importance of K\"ahler moduli backreactions once these models are viewed from a string theory perspective. In plateau-like inflation it turns out that stabilizing all moduli consistently often ruins the exponential flatness of the inflaton potential. Instead one is left with an effective potential which is exponentially steep. In natural inflation there is no such clear statement, since the periodicity of the potential usually protects these models from severe backreactions. However, there are subtleties involved when supersymmetry is broken above the Hubble scale.

%
%%
%%%
%%%%
%%%%%%%%%%%%%%%%%%%%%%%%%%%%%%%%%%%%%%%%%%%%%%%
%%%%%%%%%%%%%%%%%%%%%%%%%%%%%%%%%%%%%%%%%%%%%%%
\subsection{Starobinsky-like models without stabilizer fields}

%
%%
%%%
%%%%
%%%%%%%%%%%%%%%%%%%%%%%%%%%%%%%%%%%%%%%%%%%%%%%
\subsubsection{A model in no-scale supergravity}
\label{sec:ENO}

An appealing implementation of plateau-like inflation has been proposed by the authors of \cite{Ellis:2013xoa} and has been further developed in \cite{Ellis:2013nxa,Ellis:2013nka,Ellis:2015kqa}. The model contains two chiral superfields, denoted by $T$ and $\Phi$. It is based on the Wess-Zumino model \cite{Wess:1974tw} and its Lagrangian can be defined by
\begin{align}\label{eq:ENOK}
K &= - 3 \log \left(T + \overline T - \frac13 |\Phi|^2 \right)\,,\\
W &= M \left( \frac12 \Phi^2 - \frac{b}{3 \sqrt 3} \Phi^3 \right)\,, \label{eq:ENOW}
\end{align}
where $M$ is a mass scale which determines the energy scale of inflation. In string theory a K\"ahler potential like \eqref{eq:ENOK} can arise if $T$ is a K\"ahler modulus parameterizing the compactified volume and $\Phi$ is an untwisted matter field. If the modulus is stabilized at $T = \overline T = T_0$ one can find the scalar potential for the canonically normalized inflaton field $\varphi$ by replacing
\begin{align}
\Phi = \sqrt{6 T_0} \tanh \frac{\varphi}{\sqrt 6}\,.
\end{align} 
The corresponding imaginary part of $\Phi$ is stabilized at the origin with a Hubble-scale soft mass and decouples from inflation. The same is true for $\text{Im}\, T$. A plateau for large inflaton field values, $\varphi \gtrsim 5$, is achieved when $b = (2 T_0)^{-1/2}$, in which case the potential reads\footnote{In fact, a fine-tuning of $b$ at the level of $\mathcal O(10^{-3})$ is necessary to realize 60 $e$-folds of Starobinsky-like inflation.}
\begin{align}
V = \frac{3 M^2}{8 T_0}\left(1-e^{-\sqrt{\frac23} \varphi} \right)^2\,.
\end{align}

A crucial assumption to obtain this result is that $T$ is stabilized at its vacuum expectation value. In the following we discuss if this can be achieved with the mechanisms for K\"ahler moduli stabilization available in string theory. We distinguish two kinds of non-perturbative stabilization schemes, supersymmetric (or strong) moduli stabilization, and moduli stabilization with spontaneous supersymmetry breaking. As it turns out, unsurmountable obstacles arise in both cases.

%
%%
%%%
%%%%
%%%%%%%%%%%%%%%%%%%%%%%%%%%%%%%%%%%%%%%%%%%%%%%
\subsubsection{Strong moduli stabilization}
\label{sec:ENOstrong}

Let us assume, for the sake of simplicity, that the vacuum of our theory is such that $\alpha'$ and string-loop corrections to the K\"ahler potential are negligible in the low-energy effective description for $T$. In this case the theory is correctly described by the tree-level K\"ahler potential in Eq.~\eqref{eq:ENOK} and $T$ can be stabilized by non-perturbative contributions to $W$. In particular, we assume that 
\begin{align}\label{eq:ENOW2}
W = M \left( \frac12 \Phi^2 - \frac{b}{3 \sqrt 3} \Phi^3 \right) + W_\text{mod}(T)\,,
\end{align}
i.e., inflaton and modulus only couple gravitationally and via kinetic mixing. For $T$ to decouple from inflation $W_\text{mod}$ must be such that $m_T > H \sim M$. One may expect that the simplest way of decoupling the modulus is by supersymmetric stabilization, meaning that 
\begin{align}
W_\text{mod}(T_0) = \partial_T W_\text{mod}(T_0) = 0\,,
\end{align}
so that $F_T\big|_{T_0} = 0$, but still $\partial_T^2 W_\text{mod}(T_0) \propto m_T > H$. This has been shown to be possible via a racetrack setup \cite{Kallosh:2004yh} or via interactions with additional fields in the presence of anomalous U(1) symmetries \cite{Wieck:2014xxa}. The inflationary backreaction of moduli stabilized in this way has been previously studied in \cite{Buchmuller:2014vda} for arbitrary inflaton superpotentials. Indeed it was found that, after integrating out $T$ above the Hubble scale, the effective inflaton potential is identical to the theory without a modulus up to corrections suppressed by powers of $H/m_T$. In this case, however, this is undesirable because in the setup of \cite{Ellis:2013xoa} the no-scale symmetry associated with $T$ is crucial to realizing the plateau for large inflaton field values. If $T$ does not break supersymmetry and is decoupled with a large mass, there is no no-scale cancellation to eliminate the dangerous term proportional to $-3|W|^2$ in the scalar potential. Let us be more specific. We can compute the scalar potential defined by Eqs.~\eqref{eq:ENOK} and \eqref{eq:ENOW2} and expand around the true vacuum, $T = T_0 + \delta T(\varphi)$. By minimizing the result with respect to $\delta T(\varphi)$ we find\footnote{Without loss of generality we choose all superpotential parameters to be real. In this case only $\text{Re}\, T$ is affected by inflation, so that $\delta T (\varphi)$ is real.} 
\begin{align}
\delta T(\varphi) = \frac{3 M \tanh^2{\frac{\varphi}{\sqrt 6}}}{2\, W_\text{mod}'' (T_0)} \sim \frac{H}{m_T}\,,
\end{align}
where primes denote derivatives with respect to $T$. During inflation the modulus minimum is shifted by an amount which depends on the inflaton field value, as for the stabilizer field in Sec.~3. Inserting this result for $T$ in the potential yields the following effective potential for the inflaton,
\begin{align}\label{eq:ENOeffpot}
V = \frac{3 M^2}{8 T_0}\left(1-e^{-\sqrt{\frac23} \varphi} \right)^2 - \frac{3 M^2}{8 T_0} \sinh^4{\frac{\varphi}{\sqrt 6}} + \mathcal O\left(\frac{M}{m_T}\right)\,.
\end{align}
Hence, we obtain the Starobinsky potential but also the term proportional to $-3|W^2|$ which is negative and too steep to allow for inflation in any field range. The latter term is absent in \cite{Ellis:2013xoa} because there $T$ is taken to be a dynamical field whose vacuum expectation value is fixed in a very special way, as we discuss in Sec.~\ref{sec:AltStab}. The correction terms contained in the third piece are never large enough to remedy the model. Furthermore, Eq.~\eqref{eq:ENOeffpot} again implies $b = (2 T_0)^{-1/2}$ to high accuracy. It is a straight-forward exercise to verify that inflation is impossible for any value of $b$.

%
%%
%%%
%%%%
%%%%%%%%%%%%%%%%%%%%%%%%%%%%%%%%%%%%%%%%%%%%%%%
\subsubsection{Moduli stabilization with supersymmetry breaking}
\label{sec:ENOKKLT}

Taking this into account, one could expect that the model may be viable when $T$ breaks supersymmetry at a high scale, so that a no-scale cancellation between $F_T^2$ and $-3 |W|^2$ can take place. The interplay between such moduli stabilization schemes and large-field inflation has recently been studied in \cite{Buchmuller:2015oma}. Examples include KKLT stabilization \cite{Kachru:2003aw}, K\"ahler Uplifting \cite{Balasubramanian:2004uy,vonGersdorff:2005bf,Berg:2005yu,Westphal:2006tn}, and the Large Volume Scenario \cite{Balasubramanian:2005zx,Conlon:2005ki}. However, as pointed out in \cite{Buchmuller:2015oma}, if the modulus breaks supersymmetry the interplay between $T$ and $\Phi$ becomes non-trivial due to the appearance of soft terms, as discussed in Sec.~2. Thus, the backreaction of $T$ never fully decouples. To see this explicitly, let us choose
\begin{align}
W_\text{mod}(T) = W_0 + A e^{- a T}\,,
\end{align}
as in the mechanism of KKLT. Since the KKLT mechanism generically produces AdS vacua we include an uplift sector comprised of a Polonyi superfield $X$, so that the combined theory is defined by
\begin{align}
K &= - 3 \log{\left( T + \overline T - \frac13 |\Phi|^2 + k(|X|) \right)}\,, \\
W &= M \left( \frac12 \Phi^2 - \frac{b}{3 \sqrt 3} \Phi^3 \right) + W_\text{mod}(T) + f X\,.
\end{align}
Here $f$ denotes the scale of supersymmetry breaking and, as in Sec.~\ref{sec:Stabilizer}, the function $k$ is chosen such that $X$ is stabilized near the origin with a large mass. 

Imposing the existence of a Minkowski vacuum after inflation, at $\Phi = 0$ and $T = T_0$, leads to two relations among the parameters,
\begin{align}
A = - \frac{3 \, e^{a T_0} \, W_0}{5 + 2 a T_0}\,, \qquad f = \frac{\sqrt{12a + 6a^2 T_0}}{5 + 2 aT_0} W_0\,.
\end{align}
In this vacuum both $X$ and $T$ contribute to supersymmetry breaking, cf.~\cite{Buchmuller:2015oma} for more details. 

As in the supersymmetric case we can expand the potential around the vacuum, i.e., we expand in the displacement $\delta T (\varphi)$. In this case we find
\begin{align}
\delta T(\varphi) = - \frac{M T_0 \tanh^2{\frac{\varphi}{\sqrt 6}}}{a W_0} \sim \frac{H}{m_T}\,,
\end{align}
at leading order in $M$ and $(a T_0)^{-1}$. This leads to the following effective potential after integrating out $T$,
\begin{align}\label{eq:ENOeffpot2}
V = \frac{3 M^2}{8 T_0}\left(1-e^{-\sqrt{\frac23} \varphi} \right)^2 + \frac{3 M m_{3/2}}{\sqrt{8 T_0}} \sinh^2{\sqrt\frac23 \varphi} - \frac{3 M^2}{8 T_0} \sinh^4{\frac{\varphi}{\sqrt 6}} + \mathcal O\left(\frac{M}{m_T}\right)\,,
\end{align}
with $m_{3/2} \approx W_0/(2 T_0)^{3/2}$. As before we obtain the Starobinsky potential and the dangerous term proportional to $-3|W|^2$. In addition, there is a soft term proportional to the gravitino mass or, equivalently, to the modulus mass. In the setup of chaotic inflation studied in \cite{Buchmuller:2015oma} this soft term may dominate over the negative third term for a sufficient field range and drive inflation. The case considered here, however, seems a lost cause. Due to the steepness of the potential induced by the second term, it is never possible to achieve 60 $e$-folds of slow-roll inflation. Either the third term becomes dominant at a small field value, or the modulus is destabilized by the vacuum energy of the inflaton, rendering our four-dimensional description invalid.

Let us comment briefly on the appearance of the third term in Eq.~\eqref{eq:ENOeffpot2} and other stabilization mechanisms. Contrary to naive expectation, the F-term of $T$,
\begin{align}
F_T = e^{K/2} \sqrt{K^{T \overline T}} D_T W\Big|_{T_0} \approx - \frac{3 \sqrt3 W_0}{a (2T_0)^{5/2}}\,,
\end{align}
does not cancel the negative term by virtue of the no-scale symmetry. This is because ${F_T \sim F_X/ aT_0}$ and hence $X$ contributes dominantly to supersymmetry breaking. This situation is different in other stabilization schemes like K\"ahler Uplifting or the simplest Large Volume Scenario. In those setups the negative third term in Eq.~\eqref{eq:ENOeffpot2} may be canceled at leading order in $\mathcal V^{-1}$, the inverse of the volume of the compact manifold. The same is true for the soft term. However, even in those cases it is inconceivable that the first term in the potential is dominant for a sufficient field range, cf.~the more detailed analysis in \cite{Buchmuller:2015oma}.

%
%%
%%%
%%%%
%%%%%%%%%%%%%%%%%%%%%%%%%%%%%%%%%%%%%%%%%%%%%%%
\subsubsection{Comments on other stabilization mechanisms}
\label{sec:AltStab}

A different stabilization mechanism, without using non-perturbative superpotentials, was proposed by the authors of \cite{Ellis:2013nxa} based on the mechanism in \cite{Ellis:1984bs}. It relies on the presence of strongly stabilizing terms in the K\"ahler potential, i.e.,
\begin{align}\label{eq:ENOcheat}
K = -3 \log\left[T + \overline T - \frac13 |\Phi|^2 + \frac{\left(T + \overline T - 2 T_0\right)^4+ \left(T - \overline T \right)^4}{\Lambda^2} \right]\,,
\end{align}
while the superpotential in Eq.~\eqref{eq:ENOW} is simply extended by a constant $W_0$. The additional term in $K$ stabilizes $\text{Im}\, T$ at the origin and $\text{Re} \, T$ at $T_0$. Due to the no-scale structure of $K$ the cosmological constant vanishes in the vacuum and supersymmetry is broken by the auxiliary field of $T$, with $m_{3/2} = W_0/(2 T_0)^{3/2}$. As long as $\Lambda \ll 1$ the modulus is stabilized at a high scale with $m_T \sim m_{3/2}/\Lambda$. This distinguishes this mechanism from the others. The modulus breaks supersymmetry dominantly but there is still a hierarchy between $m_T$ and $m_{3/2}$.

We can apply the same formalism as before to integrate out $T$ and find the effective inflaton potential. We find
\begin{align}
V = V_0  \left[ 1 - \frac{V_0}{m_T^2} + \mathcal O\left(\frac{H^3}{m_T^3}\right) \right] \,, \qquad V_0 = \frac{3 M^2}{8 T_0}\left(1-e^{-\sqrt{\frac23} \varphi} \right)^2\,.
\end{align}
The result is the Starobinsky potential with a series of corrections suppressed by at least $H^2/m_T^2$. In particular, the soft mass term and the term proportional to $-3|W|^2$ are absent due to the exact no-scale symmetry of the theory. As long as the backreaction terms are under control, which is necessary for consistency since $m_T > H$ is still a requirement for stability, inflation may proceed as if the modulus was fixed by hand. This puts a bound on a combination of the parameters $W_0$, $\Lambda$, $T_0$, and $M$.

From this perspective there seems to be an elegant solution to the problems encountered in Secs.~\ref{sec:ENOstrong} and \ref{sec:ENOKKLT}. However, from the perspective of string theory it is questionable if the K\"ahler potential in Eq.~\eqref{eq:ENOcheat} is realistic. Perturbative corrections to $K$ from the $\alpha'$ and $g_\text{s}$ expansions are usually expected to include smaller (and negative) powers of $T$ \cite{Cicoli:2007xp}. Non-perturbative contributions to $K$ are typically subdominant to those contributions \cite{Kaplunovsky:1994fg,Burgess:1995aa}. On the other hand, the most successful moduli stabilization schemes compatible with string theory available so far involve the breaking of the no-scale symmetry by non-perturbative superpotentials, as demonstrated in Secs.~\ref{sec:ENOstrong} and \ref{sec:ENOKKLT}. In this case the obstacles outlined above always seem to be present and challenge the setup defined by Eqs.~\eqref{eq:ENOK} and \eqref{eq:ENOW}.

%
%%
%%%
%%%%%%%%%%%%%%%%%%%%%%%%%%%%%%%%%%%%%%%%%%%%%%%%%%%
\subsubsection{The Goncharov-Linde model}
\label{sec:goncharov}

There exist a number of other supergravity setups which feature a plateau and which are not based on no-scale supergravity. Although their string theory interpretation is rather unclear, it is instructive to consider one of the most prominent examples as a toy model. It was proposed in \cite{Goncharov:1983mw,Goncharov:1985yu} and its Lagrangian is defined by
\begin{align}\label{eq:GoLi}
K &= \frac12 (\Phi + \overline \Phi)^2 \,,\\
W &= \frac16 M \sin (\sqrt 3 \Phi) \cos (\sqrt 3 \Phi)\,,
\end{align}
in an obvious notation. The scalar potential for the canonically normalized inflaton field $\varphi = \sqrt 2 \, \text{Im} \, \Phi$ reads
\begin{align}\label{eq:GoLiPot}
V = \frac{1}{12} M^2 \left[4-\tanh^2\left( \sqrt{\frac32} \varphi\right)  \right] \tanh^2\left( \sqrt{\frac32} \varphi\right)\,.
\end{align}
It is exponentially flat for $\varphi \gg 1$. We can study the coupling of this model to supersymmetry breaking and to K\"ahler moduli in two steps. It has been noted in \cite{Linde:2014hfa} that there is an upper bound on the allowed gravitino mass once the model is coupled to a Polonyi field. Adding a Polonyi sector to Eqs.~\eqref{eq:GoLi} as in our other examples leads to an additional soft term in the scalar potential,
\begin{align}\label{eq:GoLiSoft}
V_\text{soft} = M m_{3/2} \sinh\left( \sqrt{\frac32} \varphi\right) \tanh\left( \sqrt{\frac32} \varphi\right)\,,
\end{align}
with $m_{3/2} = e^{K/2} W = W_0$. Evidently, this new term is exponentially steep and makes inflation impossible when $m_{3/2} \gtrsim 10^{-4} M \sim 10^9\, \text{GeV}$. 

A much bigger problem arises when we couple the Goncharov-Linde model to moduli with a no-scale K\"ahler potential. From the general discussion in Sec.~2 it is clear that a combined theory defined by 
\begin{align}\label{eq:GoLiKKLT}
K &= -3 \log (T + \overline T) + \frac12 (\Phi + \overline \Phi)^2 \,,\\
W &= W_\text{mod}(T) + \frac16 M \sin (\sqrt 3 \Phi) \cos (\sqrt 3 \Phi)\,,
\end{align}
will contain the soft term in Eq.~\eqref{eq:GoLiSoft} as well as correction terms in the effective potential which are proportional to the square of the inflaton-dependent piece of the superpotential, cf.~Eq.~\eqref{g17}.\footnote{This is true unless $T$ does not break supersymmetry in the ground state, as in strong moduli stabilization.} For example, for $W_\text{mod} = W_0 + A e^{-a T}$, the leading-order backreaction term reads
\begin{align}
V_\text{back} = - \frac{1}{4 a T_0} M^2 \sinh^2\left( \sqrt{\frac32} \varphi\right) \tanh^2\left( \sqrt{\frac32} \varphi\right) + \dots\,.
\end{align} 
Note that we have rescaled the mass scale $M$ by a factor of $(2 T_0)^{-3/2}$. This piece of the backreaction of $T$ is independent of the gravitino mass. Eq.~\eqref{g17} contains additional terms suppressed by powers of $H/m_T$, where $m_T \sim m_{3/2}$ denotes the mass of modulus, which produce subdominant effects.

The combined theory described by 
\begin{align}
V = V_0 + V_\text{soft} + V_\text{back}\,,
\end{align}
where $V_0$ denotes the original potential in Eq.~\eqref{eq:GoLiPot}, has two problems. First, $V_\text{back}$ is negative and exponentially steep, so that 60 $e$-folds of inflation are impossible to achieve. Second, even if the backreaction could be suppressed by some mechanism the steep soft term is dominant as long as $T$ remains stabilized, since stability requires $m_{3/2} > H$ as discussed earlier. This can be avoided, of course, if we resort to strong moduli stabilization. In that case the soft term can be made very small and all correction terms can be suppressed by making $T$ very heavy.

%
%%
%%%
%%%%%%%%%%%%%%%%%%%%%%%%%%%%%%%%%%%%%%%%%%%%%%%%%%%
%%%%%%%%%%%%%%%%%%%%%%%%%%%%%%%%%%%%%%%%%%%%%%%%%%%
\subsection{Natural inflation models without stabilizer fields}
\label{sec:NatNoStabilizer}

A class of natural inflation models without the need for a stabilizer field was proposed in \cite{Czerny:2014xja} and developed further in \cite{Czerny:2014qqa}. It is closely related to the extended setup discussed in Sec.~\ref{sec:NatStab1}. We can write it as
\begin{align}
K &= \frac12 (\Phi + \overline \Phi)^2\,,\\
W &= W_0 +  A e^{-a \Phi} + B e^{-b \Phi}\,,
\end{align}
where $\Phi$ once more contains the canonically normalized axion $\varphi$ which is protected by a shift symmetry. The above superpotential could clearly arise from a string theory compactification with $\Phi$ being a modulus field and $W_0$ resulting from fluxes and/or vacuum expectation values of heavy fields. In any case, the corresponding scalar potential for $\varphi$ reads
\begin{align}\label{eq:NatPot4}
V = \Lambda^4_0 + 2 A B (-3 + ab) \cos \left( \frac{(a-b)\varphi}{\sqrt 2} \right) - 6 A W_0 \cos \left( \frac{a\varphi}{\sqrt 2} \right) - 6 B W_0 \cos \left( \frac{b \varphi}{\sqrt 2} \right)\,,
\end{align}
where $\Lambda_0$ denotes a constant which depends on the superpotential parameters. Cancellation of the cosmological constant in the true vacuum must be ensured by an appropriate uplift, for example, via a Polonyi field.\footnote{Note that in this case there is no stabilizer field whose backreaction we have to fear. Thus, the uplift is somewhat trivial in this setup, even for large values of $W_0$.} For now, we assume $\sqrt 2 \, \text{Re} \, \Phi = \chi$ to be stabilized during inflation and in the vacuum. 

There are two interesting cases which deserve our attention. The authors of \cite{Czerny:2014xja} assumed the hierarchy $|A|, |B| \ll |W_0|$ so that $\chi$ is sufficiently heavy to decouple from the inflationary dynamics. In that case the last two terms in Eq.~\eqref{eq:NatPot4} drive inflation and the trajectory follows a modulated cosine potential which may predict observables in accordance with observations.

However, a sufficient mass splitting between $\varphi$ and $\chi$ may also be achieved when $|W_0| \ll |A|, |B|$ and at the same time $(a-b)\ll 1$. This case resembles the discussion following Eq.~\eqref{eq:NatPot2}. In this case, the second term in Eq.~\eqref{eq:NatPot4} drives inflation and the Hubble scale is $H^2 \sim A B$. The inflaton mass, on the other hand, is approximately ${m_\varphi^2 \sim AB (a-b)^2 \ll H^2}$. The resulting inflaton potential is a cosine function with an effectively large axion decay constant. The real part of $\Phi$, not being protected by the shift symmetry, receives a Hubble-scale soft mass during inflation. This is sufficient to decouple it from the inflationary dynamics.

In view of a possible string theory embedding of this model it is conceivable that the latter case with a small value of $W_0$ is difficult to reconcile with moduli stabilization. Any moduli stabilization scheme which entails supersymmetry breaking, meaning anything but strong moduli stabilization, requires $m_{3/2} \sim W_0 > H$. This is at odds with the requirement that the second term in Eq.~\eqref{eq:NatPot4} is the dominant one. Thus, one is either forced to work in a regime where $W_0$ and thus the supersymmetry breaking scale is large, and the inflaton potential is a rather complicated modulated cosine function, or to resort to strong moduli stabilization.

%
%%
%%%
%%%%
%%%%%%%%%%%%%%%%%%%%%%%%%%%%%%%%%%%%%%%%%%%%%%%
%%%%%%%%%%%%%%%%%%%%%%%%%%%%%%%%%%%%%%%%%%%%%%%
%%%%%%%%%%%%%%				   %%%%%%%%%%%%%%%%%%%%%%%
%%%%%%%%%%%%%%         Section 5       %%%%%%%%%%%%%%%%%%%%%%%
%%%%%%%%%%%%%%				   %%%%%%%%%%%%%%%%%%%%%%%
%%%%%%%%%%%%%%%%%%%%%%%%%%%%%%%%%%%%%%%%%%%%%%%
%%%%%%%%%%%%%%%%%%%%%%%%%%%%%%%%%%%%%%%%%%%%%%%

\section{Discussion and conclusion}

We have analyzed the effects of heavy stabilizer fields and heavy moduli fields on supergravity models of inflation. In the presence of supersymmetry breaking, integrating out these heavy modes generically induces backreactions which are difficult to decouple even when the masses are taken to be very large. Some effects even increase with the mass of the heavy fields. The results presented here are generalizations of \cite{Buchmuller:2014pla} and \cite{Buchmuller:2015oma} concerning the stabilizer backreaction and the moduli backreaction, respectively.

In our examples of plateau models with stabilizer fields we have found that the backreaction constrains the gravitino mass in the vacuum to be $m_{3/2} \lesssim H \sim 10^{13} \, \text{GeV}$, or even $m_{3/2} \lesssim 10^{10}\, \text{GeV}$ in cases where the correction terms affect the plateau regime of the inflaton potential. With regard to string theory embeddings of such models this implies that remnant moduli fields can not be stabilized with non-perturbative superpotentials and spontaneous supersymmetry breaking at a high scale, like in KKLT or the Large Volume Scenario. In plateau models without stabilizer fields the backreaction of moduli can have similar effects. The Goncharov-Linde model seems incompatible with anything but strong moduli stabilization. The no-scale models discussed in Sec.~\ref{sec:ENO} are even incompatible with any kind of moduli stabilization involving non-perturbative superpotentials which break the no-scale symmetry of the K\"ahler potential. In our opinion, however, these tensions put pressure on the available mechanisms of moduli stabilization rather than on inflation.

In natural inflation the picture is somewhat different. The models involving stabilizer fields are often protected from severe effects since the backreaction terms are of the same functional form as the original cosine potentials. A proper uplift can compensate the sign difference in the effective potential even for very large gravitino masses. However, this is no longer true when the periodicity of the potential is lifted by monomial functions as in setups of axion monodromy inflation. In those cases similar bounds on $m_{3/2}$ as in plateau-like inflation or chaotic inflation apply. Thus, barring a few exceptions, this confirms our previous finding that stabilizer fields do not fare well with high-energy supersymmetry. As is well-known, natural inflation without stabilizer fields and with a large effective axion decay constant is difficult to achieve without fine-tuning. We have analyzed one possible setup and found that, similar as in the Goncharov-Linde model, the scale of supersymmetry breaking must be low for the correct term to dominate the inflationary vacuum energy. This, however, is at odds with moduli stabilization unless we resort to strong moduli stabilization once more. 

Although our general discussion in Sec.~2 covers many more possible setups which exist in the literature, we could not cover all of them in explicit examples. For example, it has been realized in a series of recent publications that some of our plateau toy models belong to a more general class of so-called $\alpha$-attractor setups \cite{Kallosh:2013yoa,Kallosh:2014rga,Galante:2014ifa,Kallosh:2015lwa,Roest:2015qya,Carrasco:2015rva,Scalisi:2015qga,Carrasco:2015pla}. It may be worthwhile to study whether our results apply to these more general supergravity models as well. Furthermore, it may be interesting to investigate backreactions in, for example, the models developed in \cite{Ketov:2014qha,Ketov:2014hya}, the string theory embedding of natural inflation in \cite{Hebecker:2015rya}, and the string-effective models of \cite{Palti:2014kza,Marchesano:2014mla,Grimm:2014vva,Ibanez:2014kia,Blumenhagen:2014nba,Ibanez:2014swa,Blumenhagen:2015kja}. The setup recently proposed in \cite{Escobar:2015fda} is of particular interest since it involves both K\"ahler moduli and a stabilizer field.

%
%%
%%%
%%%%
%%%%%%%%%%%%%%%%%%%%%%%%%%%%%%%%%%%%%%%%%%%%%%%
%%%%%%%%%%%%%%%%%%%%%%%%%%%%%%%%%%%%%%%%%%%%%%%

\subsection*{Acknowledgments}
The authors would like to thank Lucien Heurtier, Andrei Linde, Francisco Pedro, Marco Scalisi, Takahiro Terada, Yvette Welling, and Alexander Westphal for stimulating discussions. C.W. would like to thank the Joachim Herz Foundation for financial support and the Instituto de Fisica Teorica (IFT UAM-CSIC) in Madrid for hospitality. E.D. thanks the Alexander von Humboldt foundation and DESY Hamburg for support and hospitality.

%
%%
%%%
%%%%
%%%%%%%%%%%%%%%%%%%%%%%%%%%%%%%%%%%%%%%%%%%%%%%
%%%%%%%%%%%%%%%%%%%%%%%%%%%%%%%%%%%%%%%%%%%%%%%


\begin{thebibliography}{99}

%\cite{Ade:2015tva}
\bibitem{Ade:2015tva} 
  P.~A.~R.~Ade {\it et al.} [BICEP2 and Planck Collaborations],
  %``Joint Analysis of BICEP2/$Keck ?Array$ and $Planck$ Data,''
  Phys.\ Rev.\ Lett.\  {\bf 114}, 101301 (2015)
  [arXiv:1502.00612 [astro-ph.CO]].
  %%CITATION = ARXIV:1502.00612;%%
  %196 citations counted in INSPIRE as of 15 sept. 2015


%\cite{Ade:2015lrj}
\bibitem{Ade:2015lrj} 
  P.~A.~R.~Ade {\it et al.} [Planck Collaboration],
  %``Planck 2015 results. XX. Constraints on inflation,''
  arXiv:1502.02114 [astro-ph.CO].
  %%CITATION = ARXIV:1502.02114;%%
  %284 citations counted in INSPIRE as of 15 sept. 2015


%\cite{Baumann:2014nda}
\bibitem{Baumann:2014nda} 
  D.~Baumann and L.~McAllister,
  %``Inflation and String Theory,''
  arXiv:1404.2601 [hep-th].
  %%CITATION = ARXIV:1404.2601;%%
  %122 citations counted in INSPIRE as of 15 sept. 2015


%\cite{Starobinsky:1980te}
\bibitem{Starobinsky:1980te} 
  A.~A.~Starobinsky,
  %``A New Type of Isotropic Cosmological Models Without Singularity,''
  Phys.\ Lett.\ B {\bf 91}, 99 (1980).
  %%CITATION = PHLTA,B91,99;%%
  %2408 citations counted in INSPIRE as of 15 sept. 2015


%\cite{Cecotti:1987sa}
\bibitem{Cecotti:1987sa} 
  S.~Cecotti,
  %``Higher Derivative Supergravity Is Equivalent To Standard Supergravity Coupled To Matter. 1.,''
  Phys.\ Lett.\ B {\bf 190}, 86 (1987).
  %%CITATION = PHLTA,B190,86;%%
  %91 citations counted in INSPIRE as of 15 sept. 2015


%\cite{Kallosh:2013lkr}
\bibitem{Kallosh:2013lkr} 
  R.~Kallosh and A.~Linde,
  %``Superconformal generalizations of the Starobinsky model,''
  JCAP {\bf 1306}, 028 (2013)
  [arXiv:1306.3214 [hep-th]].
  %%CITATION = ARXIV:1306.3214;%%
  %121 citations counted in INSPIRE as of 15 sept. 2015


%\cite{Ellis:2013xoa}
\bibitem{Ellis:2013xoa} 
  J.~Ellis, D.~V.~Nanopoulos and K.~A.~Olive,
  %``No-Scale Supergravity Realization of the Starobinsky Model of Inflation,''
  Phys.\ Rev.\ Lett.\  {\bf 111}, 111301 (2013)
  [Phys.\ Rev.\ Lett.\  {\bf 111}, no. 12, 129902 (2013)]
  [arXiv:1305.1247 [hep-th]].
  %%CITATION = ARXIV:1305.1247;%%
  %117 citations counted in INSPIRE as of 15 sept. 2015


%\cite{Goncharov:1983mw}
\bibitem{Goncharov:1983mw} 
  A.~B.~Goncharov and A.~D.~Linde,
  %``Chaotic Inflation in Supergravity,''
  Phys.\ Lett.\ B {\bf 139}, 27 (1984).
  %%CITATION = PHLTA,B139,27;%%
  %100 citations counted in INSPIRE as of 15 sept. 2015


%\cite{Goncharov:1985yu}
\bibitem{Goncharov:1985yu} 
  A.~S.~Goncharov and A.~D.~Linde,
  %``Chaotic Inflation Of The Universe In Supergravity,''
  Sov.\ Phys.\ JETP {\bf 59}, 930 (1984)
  [Zh.\ Eksp.\ Teor.\ Fiz.\  {\bf 86}, 1594 (1984)].
  %%CITATION = SPHJA,59,930;%%
  %26 citations counted in INSPIRE as of 15 sept. 2015


%\cite{Freese:1990rb}
\bibitem{Freese:1990rb} 
  K.~Freese, J.~A.~Frieman and A.~V.~Olinto,
  %``Natural inflation with pseudo - Nambu-Goldstone bosons,''
  Phys.\ Rev.\ Lett.\  {\bf 65}, 3233 (1990).
  %%CITATION = PRLTA,65,3233;%%
  %615 citations counted in INSPIRE as of 15 sept. 2015


%\cite{Banks:2003sx}
\bibitem{Banks:2003sx} 
  T.~Banks, M.~Dine, P.~J.~Fox and E.~Gorbatov,
  %``On the possibility of large axion decay constants,''
  JCAP {\bf 0306}, 001 (2003)
  [hep-th/0303252].
  %%CITATION = HEP-TH/0303252;%%
  %156 citations counted in INSPIRE as of 15 sept. 2015


%\cite{Svrcek:2006yi}
\bibitem{Svrcek:2006yi} 
  P.~Svrcek and E.~Witten,
  %``Axions In String Theory,''
  JHEP {\bf 0606}, 051 (2006)
  [hep-th/0605206].
  %%CITATION = HEP-TH/0605206;%%
  %332 citations counted in INSPIRE as of 15 sept. 2015


%\cite{Kim:2004rp}
\bibitem{Kim:2004rp} 
  J.~E.~Kim, H.~P.~Nilles and M.~Peloso,
  %``Completing natural inflation,''
  JCAP {\bf 0501}, 005 (2005)
  [hep-ph/0409138].
  %%CITATION = HEP-PH/0409138;%%
  %194 citations counted in INSPIRE as of 15 sept. 2015


%\cite{Silverstein:2008sg}
\bibitem{Silverstein:2008sg} 
  E.~Silverstein and A.~Westphal,
  %``Monodromy in the CMB: Gravity Waves and String Inflation,''
  Phys.\ Rev.\ D {\bf 78}, 106003 (2008)
  [arXiv:0803.3085 [hep-th]].
  %%CITATION = ARXIV:0803.3085;%%
  %362 citations counted in INSPIRE as of 15 sept. 2015


%\cite{McAllister:2008hb}
\bibitem{McAllister:2008hb} 
  L.~McAllister, E.~Silverstein and A.~Westphal,
  %``Gravity Waves and Linear Inflation from Axion Monodromy,''
  Phys.\ Rev.\ D {\bf 82}, 046003 (2010)
  [arXiv:0808.0706 [hep-th]].
  %%CITATION = ARXIV:0808.0706;%%
  %337 citations counted in INSPIRE as of 15 sept. 2015


%\cite{Kallosh:2004yh}
\bibitem{Kallosh:2004yh} 
  R.~Kallosh and A.~D.~Linde,
  %``Landscape, the scale of SUSY breaking, and inflation,''
  JHEP {\bf 0412}, 004 (2004)
  [hep-th/0411011].
  %%CITATION = HEP-TH/0411011;%%
  %214 citations counted in INSPIRE as of 15 sept. 2015


%\cite{Dudas:2012wi}
\bibitem{Dudas:2012wi} 
  E.~Dudas, A.~Linde, Y.~Mambrini, A.~Mustafayev and K.~A.~Olive,
  %``Strong moduli stabilization and phenomenology,''
  Eur.\ Phys.\ J.\ C {\bf 73}, no. 1, 2268 (2013)
  [arXiv:1209.0499 [hep-ph]].
  %%CITATION = ARXIV:1209.0499;%%
  %53 citations counted in INSPIRE as of 15 sept. 2015


%\cite{Wieck:2014xxa}
\bibitem{Wieck:2014xxa} 
  C.~Wieck and M.~W.~Winkler,
  %``Inflation with Fayet-Iliopoulos Terms,''
  Phys.\ Rev.\ D {\bf 90}, no. 10, 103507 (2014)
  [arXiv:1408.2826 [hep-th]].
  %%CITATION = ARXIV:1408.2826;%%
  %13 citations counted in INSPIRE as of 15 sept. 2015


%\cite{Buchmuller:2014vda}
\bibitem{Buchmuller:2014vda} 
  W.~Buchmuller, C.~Wieck and M.~W.~Winkler,
  %``Supersymmetric Moduli Stabilization and High-Scale Inflation,''
  Phys.\ Lett.\ B {\bf 736}, 237 (2014)
  [arXiv:1404.2275 [hep-th]].
  %%CITATION = ARXIV:1404.2275;%%
  %18 citations counted in INSPIRE as of 15 sept. 2015


%\cite{Kachru:2003aw}
\bibitem{Kachru:2003aw} 
  S.~Kachru, R.~Kallosh, A.~D.~Linde and S.~P.~Trivedi,
  %``De Sitter vacua in string theory,''
  Phys.\ Rev.\ D {\bf 68}, 046005 (2003)
  [hep-th/0301240].
  %%CITATION = HEP-TH/0301240;%%
  %2125 citations counted in INSPIRE as of 15 sept. 2015


%\cite{Balasubramanian:2005zx}
\bibitem{Balasubramanian:2005zx} 
  V.~Balasubramanian, P.~Berglund, J.~P.~Conlon and F.~Quevedo,
  %``Systematics of moduli stabilisation in Calabi-Yau flux compactifications,''
  JHEP {\bf 0503}, 007 (2005)
  [hep-th/0502058].
  %%CITATION = HEP-TH/0502058;%%
  %506 citations counted in INSPIRE as of 15 sept. 2015


%\cite{Conlon:2005ki}
\bibitem{Conlon:2005ki} 
  J.~P.~Conlon, F.~Quevedo and K.~Suruliz,
  %``Large-volume flux compactifications: Moduli spectrum and D3/D7 soft supersymmetry breaking,''
  JHEP {\bf 0508}, 007 (2005)
  [hep-th/0505076].
  %%CITATION = HEP-TH/0505076;%%
  %293 citations counted in INSPIRE as of 15 sept. 2015


%\cite{Balasubramanian:2004uy}
\bibitem{Balasubramanian:2004uy} 
  V.~Balasubramanian and P.~Berglund,
  %``Stringy corrections to Kahler potentials, SUSY breaking, and the cosmological constant problem,''
  JHEP {\bf 0411}, 085 (2004)
  [hep-th/0408054].
  %%CITATION = HEP-TH/0408054;%%
  %146 citations counted in INSPIRE as of 15 sept. 2015


%\cite{vonGersdorff:2005bf}
\bibitem{vonGersdorff:2005bf} 
  G.~von Gersdorff and A.~Hebecker,
  %``Kahler corrections for the volume modulus of flux compactifications,''
  Phys.\ Lett.\ B {\bf 624}, 270 (2005)
  [hep-th/0507131].
  %%CITATION = HEP-TH/0507131;%%
  %82 citations counted in INSPIRE as of 15 sept. 2015


%\cite{Berg:2005yu}
\bibitem{Berg:2005yu} 
  M.~Berg, M.~Haack and B.~Kors,
  %``On volume stabilization by quantum corrections,''
  Phys.\ Rev.\ Lett.\  {\bf 96}, 021601 (2006)
  [hep-th/0508171].
  %%CITATION = HEP-TH/0508171;%%
  %92 citations counted in INSPIRE as of 15 sept. 2015


%\cite{Westphal:2006tn}
\bibitem{Westphal:2006tn} 
  A.~Westphal,
  %``de Sitter string vacua from Kahler uplifting,''
  JHEP {\bf 0703}, 102 (2007)
  [hep-th/0611332].
  %%CITATION = HEP-TH/0611332;%%
  %57 citations counted in INSPIRE as of 15 sept. 2015


%\cite{Buchmuller:2015oma}
\bibitem{Buchmuller:2015oma} 
  W.~Buchmuller, E.~Dudas, L.~Heurtier, A.~Westphal, C.~Wieck and M.~W.~Winkler,
  %``Challenges for Large-Field Inflation and Moduli Stabilization,''
  JHEP {\bf 1504}, 058 (2015)
  [arXiv:1501.05812 [hep-th]].
  %%CITATION = ARXIV:1501.05812;%%
  %17 citations counted in INSPIRE as of 15 sept. 2015


%\cite{Choudhury:2014sxa}
\bibitem{Choudhury:2014sxa} 
  S.~Choudhury, A.~Mazumdar and E.~Pukartas,
  %``Constraining ${\cal N}=1$ supergravity inflationary framework with non-minimal KŠhler operators,''
  JHEP {\bf 1404}, 077 (2014)
  [arXiv:1402.1227 [hep-th]].
  %%CITATION = ARXIV:1402.1227;%%
  %12 citations counted in INSPIRE as of 15 sept. 2015


%\cite{Buchmuller:2014pla}
\bibitem{Buchmuller:2014pla} 
  W.~Buchmuller, E.~Dudas, L.~Heurtier and C.~Wieck,
  %``Large-Field Inflation and Supersymmetry Breaking,''
  JHEP {\bf 1409}, 053 (2014)
  [arXiv:1407.0253 [hep-th]].
  %%CITATION = ARXIV:1407.0253;%%
  %18 citations counted in INSPIRE as of 15 sept. 2015


%\cite{Achucarro:2010jv}
\bibitem{Achucarro:2010jv} 
  A.~Achucarro, J.~O.~Gong, S.~Hardeman, G.~A.~Palma and S.~P.~Patil,
  %``Mass hierarchies and non-decoupling in multi-scalar field dynamics,''
  Phys.\ Rev.\ D {\bf 84}, 043502 (2011)
  [arXiv:1005.3848 [hep-th]].
  %%CITATION = ARXIV:1005.3848;%%
  %72 citations counted in INSPIRE as of 15 sept. 2015


%\cite{Achucarro:2010da}
\bibitem{Achucarro:2010da} 
  A.~Achucarro, J.~O.~Gong, S.~Hardeman, G.~A.~Palma and S.~P.~Patil,
  %``Features of heavy physics in the CMB power spectrum,''
  JCAP {\bf 1101}, 030 (2011)
  [arXiv:1010.3693 [hep-ph]].
  %%CITATION = ARXIV:1010.3693;%%
  %135 citations counted in INSPIRE as of 15 sept. 2015


%\cite{Achucarro:2012sm}
\bibitem{Achucarro:2012sm} 
  A.~Achucarro, J.~O.~Gong, S.~Hardeman, G.~A.~Palma and S.~P.~Patil,
  %``Effective theories of single field inflation when heavy fields matter,''
  JHEP {\bf 1205}, 066 (2012)
  [arXiv:1201.6342 [hep-th]].
  %%CITATION = ARXIV:1201.6342;%%
  %79 citations counted in INSPIRE as of 15 sept. 2015


%\cite{Achucarro:2012yr}
\bibitem{Achucarro:2012yr} 
  A.~Achucarro, V.~Atal, S.~Cespedes, J.~O.~Gong, G.~A.~Palma and S.~P.~Patil,
  %``Heavy fields, reduced speeds of sound and decoupling during inflation,''
  Phys.\ Rev.\ D {\bf 86}, 121301 (2012)
  [arXiv:1205.0710 [hep-th]].
  %%CITATION = ARXIV:1205.0710;%%
  %54 citations counted in INSPIRE as of 15 sept. 2015


%\cite{Achucarro:2015bra}
\bibitem{Achucarro:2015bra} 
  A.~Achœcarro and Y.~Welling,
  %``Multiple Field Inflation and Signatures of Heavy Physics in the CMB,''
  arXiv:1502.04369 [gr-qc].
  %%CITATION = ARXIV:1502.04369;%%
  %6 citations counted in INSPIRE as of 15 sept. 2015


%\cite{Achucarro:2015rfa}
\bibitem{Achucarro:2015rfa} 
  A.~Achœcarro, V.~Atal and Y.~Welling,
  %``On the viability of $m^2\phi^2$ and natural inflation,''
  JCAP {\bf 1507}, 008 (2015)
  [arXiv:1503.07486 [astro-ph.CO]].
  %%CITATION = ARXIV:1503.07486;%%
  %2 citations counted in INSPIRE as of 15 sept. 2015


%\cite{Kawasaki:2000yn}
\bibitem{Kawasaki:2000yn} 
  M.~Kawasaki, M.~Yamaguchi and T.~Yanagida,
  %``Natural chaotic inflation in supergravity,''
  Phys.\ Rev.\ Lett.\  {\bf 85}, 3572 (2000)
  [hep-ph/0004243].
  %%CITATION = HEP-PH/0004243;%%
  %303 citations counted in INSPIRE as of 15 sept. 2015


%\cite{Conlon:2005jm}
\bibitem{Conlon:2005jm} 
  J.~P.~Conlon and F.~Quevedo,
  %``Kahler moduli inflation,''
  JHEP {\bf 0601}, 146 (2006)
  [hep-th/0509012].
  %%CITATION = HEP-TH/0509012;%%
  %176 citations counted in INSPIRE as of 15 sept. 2015


%\cite{Cicoli:2008gp}
\bibitem{Cicoli:2008gp} 
  M.~Cicoli, C.~P.~Burgess and F.~Quevedo,
  %``Fibre Inflation: Observable Gravity Waves from IIB String Compactifications,''
  JCAP {\bf 0903}, 013 (2009)
  [arXiv:0808.0691 [hep-th]].
  %%CITATION = ARXIV:0808.0691;%%
  %90 citations counted in INSPIRE as of 15 sept. 2015


%\cite{Cicoli:2011ct}
\bibitem{Cicoli:2011ct} 
  M.~Cicoli, F.~G.~Pedro and G.~Tasinato,
  %``Poly-instanton Inflation,''
  JCAP {\bf 1112}, 022 (2011)
  [arXiv:1110.6182 [hep-th]].
  %%CITATION = ARXIV:1110.6182;%%
  %32 citations counted in INSPIRE as of 15 sept. 2015


%\cite{Blumenhagen:2012ue}
\bibitem{Blumenhagen:2012ue} 
  R.~Blumenhagen, X.~Gao, T.~Rahn and P.~Shukla,
  %``Moduli Stabilization and Inflationary Cosmology with Poly-Instantons in Type IIB Orientifolds,''
  JHEP {\bf 1211}, 101 (2012)
  [arXiv:1208.1160 [hep-th]].
  %%CITATION = ARXIV:1208.1160;%%
  %12 citations counted in INSPIRE as of 15 sept. 2015


%\cite{Bielleman:2015ina}
\bibitem{Bielleman:2015ina} 
  S.~Bielleman, L.~E.~Ibanez and I.~Valenzuela,
  %``Minkowski 3-forms, Flux String Vacua, Axion Stability and Naturalness,''
  arXiv:1507.06793 [hep-th].
  %%CITATION = ARXIV:1507.06793;%%


%\cite{Kaloper:2008fb}
\bibitem{Kaloper:2008fb} 
  N.~Kaloper and L.~Sorbo,
  %``A Natural Framework for Chaotic Inflation,''
  Phys.\ Rev.\ Lett.\  {\bf 102}, 121301 (2009)
  [arXiv:0811.1989 [hep-th]].
  %%CITATION = ARXIV:0811.1989;%%
  %129 citations counted in INSPIRE as of 15 sept. 2015


%\cite{Kaloper:2011jz}
\bibitem{Kaloper:2011jz} 
  N.~Kaloper, A.~Lawrence and L.~Sorbo,
  %``An Ignoble Approach to Large Field Inflation,''
  JCAP {\bf 1103}, 023 (2011)
  [arXiv:1101.0026 [hep-th]].
  %%CITATION = ARXIV:1101.0026;%%
  %126 citations counted in INSPIRE as of 15 sept. 2015


%\cite{Kaloper:2014zba}
\bibitem{Kaloper:2014zba} 
  N.~Kaloper and A.~Lawrence,
  %``Natural chaotic inflation and ultraviolet sensitivity,''
  Phys.\ Rev.\ D {\bf 90}, no. 2, 023506 (2014)
  [arXiv:1404.2912 [hep-th]].
  %%CITATION = ARXIV:1404.2912;%%
  %53 citations counted in INSPIRE as of 15 sept. 2015


%\cite{Roest:2013aoa}
\bibitem{Roest:2013aoa} 
  D.~Roest, M.~Scalisi and I.~Zavala,
  %``KŠhler potentials for Planck inflation,''
  JCAP {\bf 1311}, 007 (2013)
  [arXiv:1307.4343].
  %%CITATION = ARXIV:1307.4343;%%
  %29 citations counted in INSPIRE as of 15 sept. 2015


%\cite{Dudas:2014pva}
\bibitem{Dudas:2014pva} 
  E.~Dudas,
  %``Three-form multiplet and Inflation,''
  JHEP {\bf 1412}, 014 (2014)
  [arXiv:1407.5688 [hep-th]].
  %%CITATION = ARXIV:1407.5688;%%
  %12 citations counted in INSPIRE as of 15 sept. 2015


%\cite{Grisaru:1996ve}
\bibitem{Grisaru:1996ve} 
  M.~T.~Grisaru, M.~Rocek and R.~von Unge,
  %``Effective Kahler potentials,''
  Phys.\ Lett.\ B {\bf 383}, 415 (1996)
  [hep-th/9605149].
  %%CITATION = HEP-TH/9605149;%%
  %107 citations counted in INSPIRE as of 15 sept. 2015


%\cite{Antoniadis:2014oya}
\bibitem{Antoniadis:2014oya} 
  I.~Antoniadis, E.~Dudas, S.~Ferrara and A.~Sagnotti,
  %``The Volkov-Akulov-Starobinsky supergravity,''
  Phys.\ Lett.\ B {\bf 733}, 32 (2014)
  [arXiv:1403.3269 [hep-th]].
  %%CITATION = ARXIV:1403.3269;%%
  %59 citations counted in INSPIRE as of 15 sept. 2015


%\cite{Ferrara:2014kva}
\bibitem{Ferrara:2014kva} 
  S.~Ferrara, R.~Kallosh and A.~Linde,
  %``Cosmology with Nilpotent Superfields,''
  JHEP {\bf 1410}, 143 (2014)
  [arXiv:1408.4096 [hep-th]].
  %%CITATION = ARXIV:1408.4096;%%
  %45 citations counted in INSPIRE as of 15 sept. 2015


%\cite{Kallosh:2014via}
\bibitem{Kallosh:2014via} 
  R.~Kallosh and A.~Linde,
  %``Inflation and Uplifting with Nilpotent Superfields,''
  JCAP {\bf 1501}, no. 01, 025 (2015)
  [arXiv:1408.5950 [hep-th]].
  %%CITATION = ARXIV:1408.5950;%%
  %36 citations counted in INSPIRE as of 15 sept. 2015


%\cite{Dall'Agata:2014oka}
\bibitem{Dall'Agata:2014oka} 
  G.~Dall'Agata and F.~Zwirner,
  %``On sgoldstino-less supergravity models of inflation,''
  JHEP {\bf 1412}, 172 (2014)
  [arXiv:1411.2605 [hep-th]].
  %%CITATION = ARXIV:1411.2605;%%
  %30 citations counted in INSPIRE as of 15 sept. 2015


%\cite{Kallosh:2014vja}
\bibitem{Kallosh:2014vja} 
  R.~Kallosh, A.~Linde and B.~Vercnocke,
  %``Natural Inflation in Supergravity and Beyond,''
  Phys.\ Rev.\ D {\bf 90}, no. 4, 041303 (2014)
  [arXiv:1404.6244 [hep-th]].
  %%CITATION = ARXIV:1404.6244;%%
  %33 citations counted in INSPIRE as of 15 sept. 2015


%\cite{Kappl:2015pxa}
\bibitem{Kappl:2015pxa} 
  R.~Kappl, H.~P.~Nilles and M.~W.~Winkler,
  %``Natural Inflation and Low Energy Supersymmetry,''
  Phys.\ Lett.\ B {\bf 746}, 15 (2015)
  [arXiv:1503.01777 [hep-th]].
  %%CITATION = ARXIV:1503.01777;%%
  %9 citations counted in INSPIRE as of 15 sept. 2015


%\cite{Ruehle:2015afa}
\bibitem{Ruehle:2015afa} 
  F.~Ruehle and C.~Wieck,
  %``Natural inflation and moduli stabilization in heterotic orbifolds,''
  JHEP {\bf 1505}, 112 (2015)
  [arXiv:1503.07183 [hep-th]].
  %%CITATION = ARXIV:1503.07183;%%
  %6 citations counted in INSPIRE as of 15 sept. 2015


%\cite{Ellis:2013nxa}
\bibitem{Ellis:2013nxa} 
  J.~Ellis, D.~V.~Nanopoulos and K.~A.~Olive,
  %``Starobinsky-like Inflationary Models as Avatars of No-Scale Supergravity,''
  JCAP {\bf 1310}, 009 (2013)
  [arXiv:1307.3537 [hep-th]].
  %%CITATION = ARXIV:1307.3537;%%
  %102 citations counted in INSPIRE as of 15 sept. 2015


%\cite{Ellis:2013nka}
\bibitem{Ellis:2013nka} 
  J.~Ellis, D.~V.~Nanopoulos and K.~A.~Olive,
  %``A no-scale supergravity framework for sub-Planckian physics,''
  Phys.\ Rev.\ D {\bf 89}, no. 4, 043502 (2014)
  [arXiv:1310.4770 [hep-ph]].
  %%CITATION = ARXIV:1310.4770;%%
  %27 citations counted in INSPIRE as of 15 sept. 2015


%\cite{Ellis:2015kqa}
\bibitem{Ellis:2015kqa} 
  J.~Ellis, M.~A.~G.~Garcia, D.~V.~Nanopoulos and K.~A.~Olive,
  %``Phenomenological Aspects of No-Scale Inflation Models,''
  arXiv:1503.08867 [hep-ph].
  %%CITATION = ARXIV:1503.08867;%%
  %5 citations counted in INSPIRE as of 15 sept. 2015


%\cite{Wess:1974tw}
\bibitem{Wess:1974tw} 
  J.~Wess and B.~Zumino,
  %``Supergauge Transformations in Four-Dimensions,''
  Nucl.\ Phys.\ B {\bf 70}, 39 (1974).
  %%CITATION = NUPHA,B70,39;%%
  %2264 citations counted in INSPIRE as of 15 sept. 2015


%\cite{Ellis:1984bs}
\bibitem{Ellis:1984bs} 
  J.~R.~Ellis, C.~Kounnas and D.~V.~Nanopoulos,
  %``No Scale Supergravity Models with a Planck Mass Gravitino,''
  Phys.\ Lett.\ B {\bf 143}, 410 (1984).
  %%CITATION = PHLTA,B143,410;%%
  %137 citations counted in INSPIRE as of 15 sept. 2015


%\cite{Cicoli:2007xp}
\bibitem{Cicoli:2007xp} 
  M.~Cicoli, J.~P.~Conlon and F.~Quevedo,
  %``Systematics of String Loop Corrections in Type IIB Calabi-Yau Flux Compactifications,''
  JHEP {\bf 0801}, 052 (2008)
  [arXiv:0708.1873 [hep-th]].
  %%CITATION = ARXIV:0708.1873;%%
  %102 citations counted in INSPIRE as of 15 sept. 2015


%\cite{Kaplunovsky:1994fg}
\bibitem{Kaplunovsky:1994fg} 
  V.~Kaplunovsky and J.~Louis,
  %``Field dependent gauge couplings in locally supersymmetric effective quantum field theories,''
  Nucl.\ Phys.\ B {\bf 422}, 57 (1994)
  [hep-th/9402005].
  %%CITATION = HEP-TH/9402005;%%
  %201 citations counted in INSPIRE as of 15 sept. 2015


%\cite{Burgess:1995aa}
\bibitem{Burgess:1995aa} 
  C.~P.~Burgess, J.~P.~Derendinger, F.~Quevedo and M.~Quiros,
  %``On gaugino condensation with field dependent gauge couplings,''
  Annals Phys.\  {\bf 250}, 193 (1996)
  [hep-th/9505171].
  %%CITATION = HEP-TH/9505171;%%
  %87 citations counted in INSPIRE as of 15 sept. 2015


%\cite{Linde:2014hfa}
\bibitem{Linde:2014hfa} 
  A.~Linde,
  %``Does the first chaotic inflation model in supergravity provide the best fit to the Planck data?,''
  JCAP {\bf 1502}, no. 02, 030 (2015)
  [arXiv:1412.7111 [hep-th]].
  %%CITATION = ARXIV:1412.7111;%%
  %16 citations counted in INSPIRE as of 15 sept. 2015


%\cite{Czerny:2014xja}
\bibitem{Czerny:2014xja} 
  M.~Czerny, T.~Higaki and F.~Takahashi,
  %``Multi-Natural Inflation in Supergravity,''
  JHEP {\bf 1405}, 144 (2014)
  [arXiv:1403.0410 [hep-ph]].
  %%CITATION = ARXIV:1403.0410;%%
  %32 citations counted in INSPIRE as of 15 sept. 2015


%\cite{Czerny:2014qqa}
\bibitem{Czerny:2014qqa} 
  M.~Czerny, T.~Higaki and F.~Takahashi,
  %``Multi-Natural Inflation in Supergravity and BICEP2,''
  Phys.\ Lett.\ B {\bf 734}, 167 (2014)
  [arXiv:1403.5883 [hep-ph]].
  %%CITATION = ARXIV:1403.5883;%%
  %40 citations counted in INSPIRE as of 15 sept. 2015


%\cite{Kallosh:2013yoa}
\bibitem{Kallosh:2013yoa} 
  R.~Kallosh, A.~Linde and D.~Roest,
  %``Superconformal Inflationary $\alpha$-Attractors,''
  JHEP {\bf 1311}, 198 (2013)
  [arXiv:1311.0472 [hep-th]].
  %%CITATION = ARXIV:1311.0472;%%
  %72 citations counted in INSPIRE as of 15 sept. 2015


%\cite{Kallosh:2014rga}
\bibitem{Kallosh:2014rga} 
  R.~Kallosh, A.~Linde and D.~Roest,
  %``Large field inflation and double $\alpha$-attractors,''
  JHEP {\bf 1408}, 052 (2014)
  [arXiv:1405.3646 [hep-th]].
  %%CITATION = ARXIV:1405.3646;%%
  %34 citations counted in INSPIRE as of 15 sept. 2015


%\cite{Galante:2014ifa}
\bibitem{Galante:2014ifa} 
  M.~Galante, R.~Kallosh, A.~Linde and D.~Roest,
  %``Unity of Cosmological Inflation Attractors,''
  Phys.\ Rev.\ Lett.\  {\bf 114}, no. 14, 141302 (2015)
  [arXiv:1412.3797 [hep-th]].
  %%CITATION = ARXIV:1412.3797;%%
  %20 citations counted in INSPIRE as of 15 sept. 2015


%\cite{Kallosh:2015lwa}
\bibitem{Kallosh:2015lwa} 
  R.~Kallosh and A.~Linde,
  %``Planck, LHC, and ?-attractors,''
  Phys.\ Rev.\ D {\bf 91}, 083528 (2015)
  [arXiv:1502.07733 [astro-ph.CO]].
  %%CITATION = ARXIV:1502.07733;%%
  %16 citations counted in INSPIRE as of 15 sept. 2015


%\cite{Roest:2015qya}
\bibitem{Roest:2015qya} 
  D.~Roest and M.~Scalisi,
  %``Cosmological attractors from ?-scale supergravity,''
  Phys.\ Rev.\ D {\bf 92}, no. 4, 043525 (2015)
  [arXiv:1503.07909 [hep-th]].
  %%CITATION = ARXIV:1503.07909;%%
  %13 citations counted in INSPIRE as of 15 sept. 2015


%\cite{Carrasco:2015rva}
\bibitem{Carrasco:2015rva} 
  J.~J.~M.~Carrasco, R.~Kallosh and A.~Linde,
  %``Cosmological Attractors and Initial Conditions for Inflation,''
  arXiv:1506.00936 [hep-th].
  %%CITATION = ARXIV:1506.00936;%%
  %5 citations counted in INSPIRE as of 15 sept. 2015


%\cite{Scalisi:2015qga}
\bibitem{Scalisi:2015qga} 
  M.~Scalisi,
  %``Cosmological $\alpha$-Attractors and de Sitter Landscape,''
  arXiv:1506.01368 [hep-th].
  %%CITATION = ARXIV:1506.01368;%%
  %8 citations counted in INSPIRE as of 15 sept. 2015


%\cite{Carrasco:2015pla}
\bibitem{Carrasco:2015pla} 
  J.~J.~M.~Carrasco, R.~Kallosh and A.~Linde,
  %``$\alpha $-Attractors: Planck, LHC and Dark Energy,''
  arXiv:1506.01708 [hep-th].
  %%CITATION = ARXIV:1506.01708;%%
  %10 citations counted in INSPIRE as of 15 sept. 2015


%\cite{Ketov:2014qha}
\bibitem{Ketov:2014qha} 
  S.~V.~Ketov and T.~Terada,
  %``Inflation in Supergravity with a Single Chiral Superfield,''
  Phys.\ Lett.\ B {\bf 736}, 272 (2014)
  [arXiv:1406.0252 [hep-th]].
  %%CITATION = ARXIV:1406.0252;%%
  %24 citations counted in INSPIRE as of 15 sept. 2015


%\cite{Ketov:2014hya}
\bibitem{Ketov:2014hya} 
  S.~V.~Ketov and T.~Terada,
  %``Generic Scalar Potentials for Inflation in Supergravity with a Single Chiral Superfield,''
  JHEP {\bf 1412}, 062 (2014)
  [arXiv:1408.6524 [hep-th]].
  %%CITATION = ARXIV:1408.6524;%%
  %23 citations counted in INSPIRE as of 15 sept. 2015


%\cite{Hebecker:2015rya}
\bibitem{Hebecker:2015rya} 
  A.~Hebecker, P.~Mangat, F.~Rompineve and L.~T.~Witkowski,
  %``Winding out of the Swamp: Evading the Weak Gravity Conjecture with F-term Winding Inflation?,''
  Phys.\ Lett.\ B {\bf 748}, 455 (2015)
  [arXiv:1503.07912 [hep-th]].
  %%CITATION = ARXIV:1503.07912;%%
  %10 citations counted in INSPIRE as of 15 sept. 2015


%\cite{Palti:2014kza}
\bibitem{Palti:2014kza} 
  E.~Palti and T.~Weigand,
  %``Towards large r from [p, q]-inflation,''
  JHEP {\bf 1404}, 155 (2014)
  [arXiv:1403.7507 [hep-th]].
  %%CITATION = ARXIV:1403.7507;%%
  %50 citations counted in INSPIRE as of 15 sept. 2015


%\cite{Marchesano:2014mla}
\bibitem{Marchesano:2014mla} 
  F.~Marchesano, G.~Shiu and A.~M.~Uranga,
  %``F-term Axion Monodromy Inflation,''
  JHEP {\bf 1409}, 184 (2014)
  [arXiv:1404.3040 [hep-th]].
  %%CITATION = ARXIV:1404.3040;%%
  %75 citations counted in INSPIRE as of 15 sept. 2015


%\cite{Grimm:2014vva}
\bibitem{Grimm:2014vva} 
  T.~W.~Grimm,
  %``Axion Inflation in F-theory,''
  Phys.\ Lett.\ B {\bf 739}, 201 (2014)
  [arXiv:1404.4268 [hep-th]].
  %%CITATION = ARXIV:1404.4268;%%
  %41 citations counted in INSPIRE as of 15 sept. 2015


%\cite{Ibanez:2014kia}
\bibitem{Ibanez:2014kia} 
  L.~E.~Ibanez and I.~Valenzuela,
  %``The inflaton as an MSSM Higgs and open string modulus monodromy inflation,''
  Phys.\ Lett.\ B {\bf 736}, 226 (2014)
  [arXiv:1404.5235 [hep-th]].
  %%CITATION = ARXIV:1404.5235;%%
  %33 citations counted in INSPIRE as of 15 sept. 2015


%\cite{Blumenhagen:2014nba}
\bibitem{Blumenhagen:2014nba} 
  R.~Blumenhagen, D.~Herschmann and E.~Plauschinn,
  %``The Challenge of Realizing F-term Axion Monodromy Inflation in String Theory,''
  JHEP {\bf 1501}, 007 (2015)
  [arXiv:1409.7075 [hep-th]].
  %%CITATION = ARXIV:1409.7075;%%
  %30 citations counted in INSPIRE as of 15 sept. 2015


%\cite{Ibanez:2014swa}
\bibitem{Ibanez:2014swa} 
  L.~E.~Ibanez, F.~Marchesano and I.~Valenzuela,
  %``Higgs-otic Inflation and String Theory,''
  JHEP {\bf 1501}, 128 (2015)
  [arXiv:1411.5380 [hep-th]].
  %%CITATION = ARXIV:1411.5380;%%
  %21 citations counted in INSPIRE as of 15 sept. 2015


%\cite{Blumenhagen:2015kja}
\bibitem{Blumenhagen:2015kja} 
  R.~Blumenhagen, A.~Font, M.~Fuchs, D.~Herschmann, E.~Plauschinn, Y.~Sekiguchi and F.~Wolf,
  %``A Flux-Scaling Scenario for High-Scale Moduli Stabilization in String Theory,''
  Nucl.\ Phys.\ B {\bf 897}, 500 (2015)
  [arXiv:1503.07634 [hep-th]].
  %%CITATION = ARXIV:1503.07634;%%
  %10 citations counted in INSPIRE as of 15 sept. 2015


%\cite{Escobar:2015fda}
\bibitem{Escobar:2015fda} 
  D.~Escobar, A.~Landete, F.~Marchesano and D.~Regalado,
  %``Large field inflation from D-branes,''
  arXiv:1505.07871 [hep-th].
  %%CITATION = ARXIV:1505.07871;%%
  %1 citations counted in INSPIRE as of 15 sept. 2015


\end{thebibliography}
\end{document}